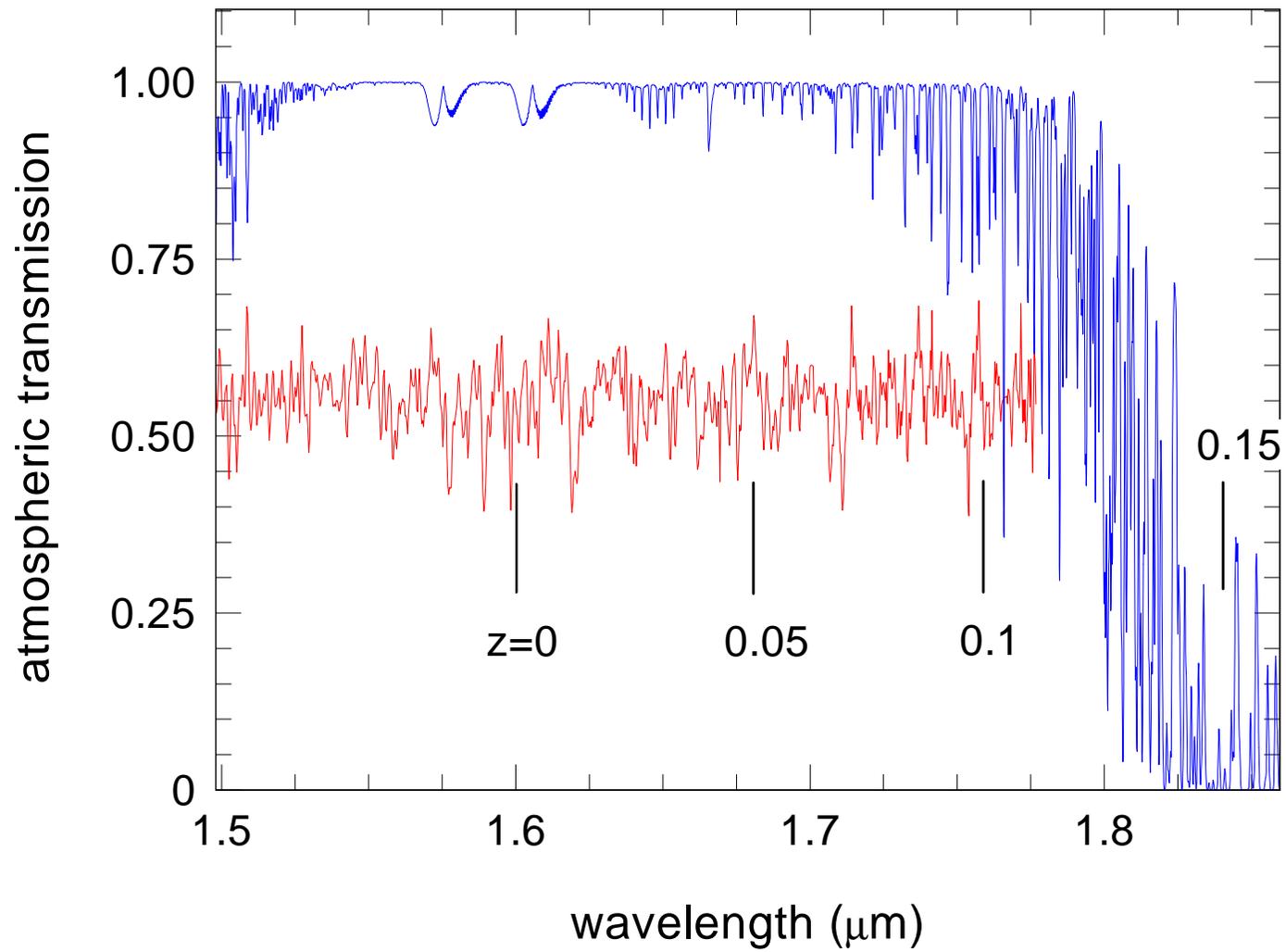

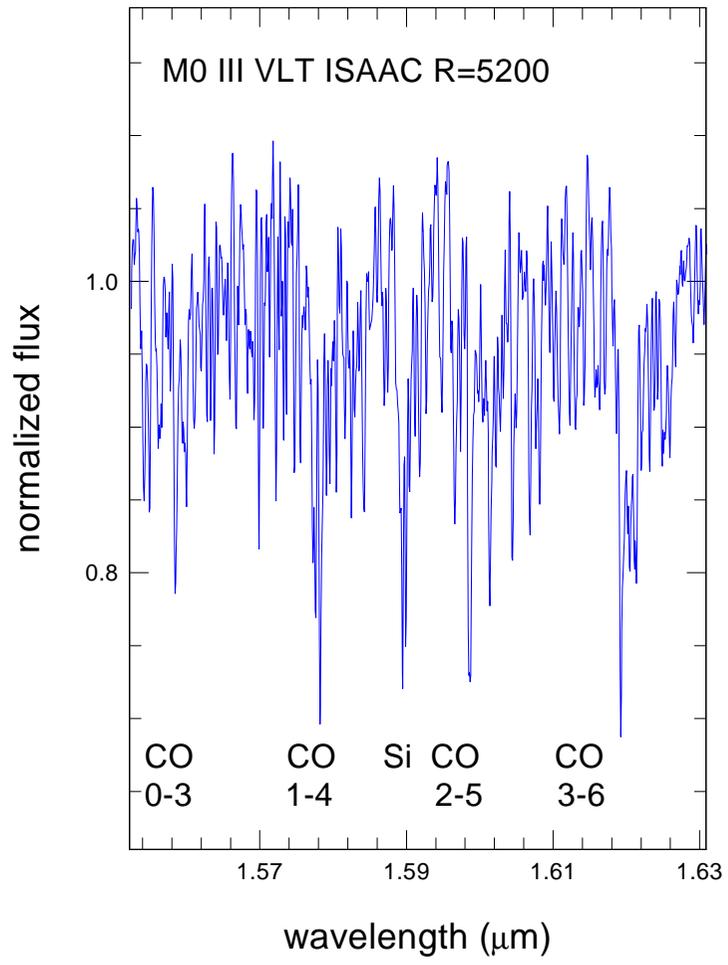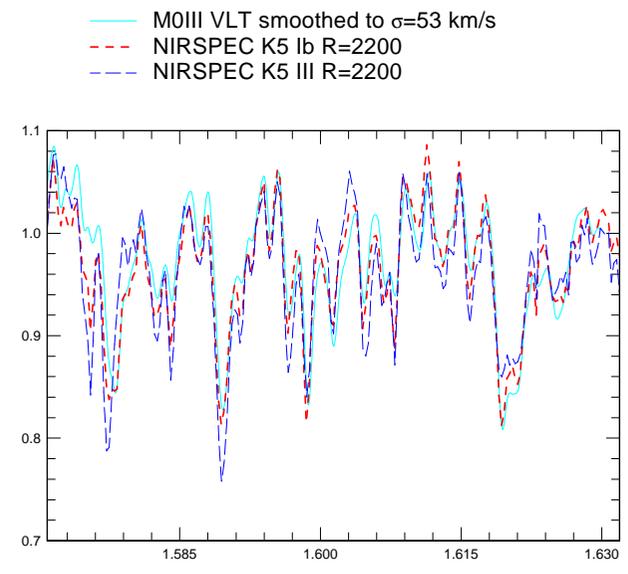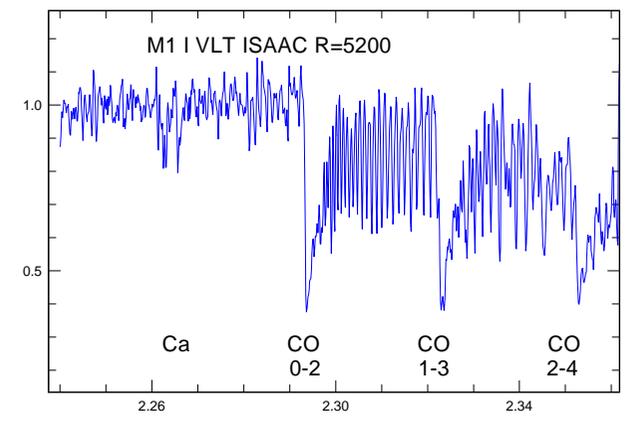

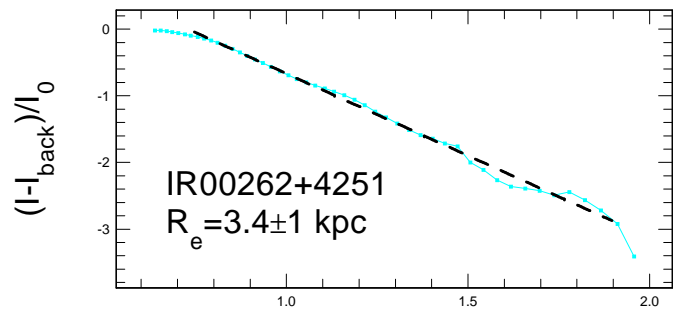
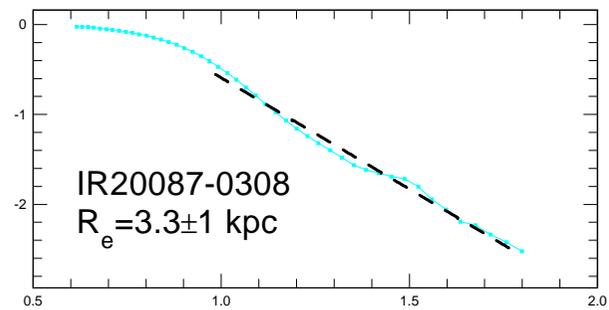
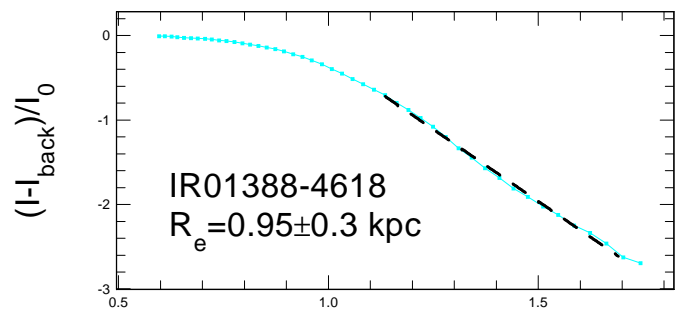
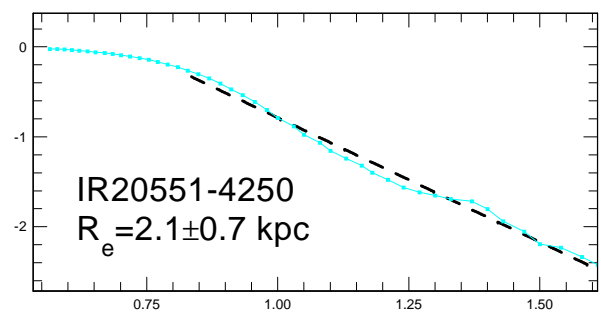
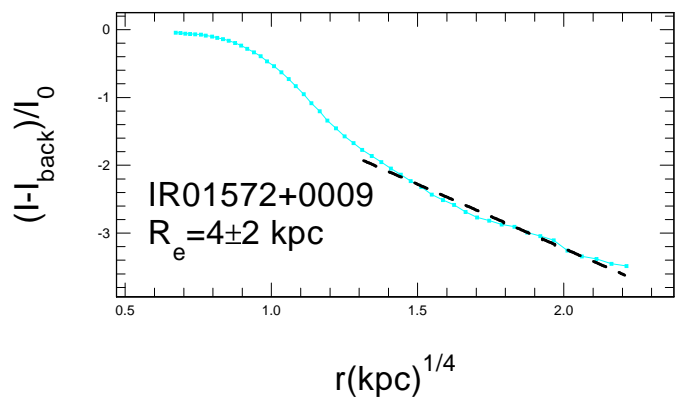
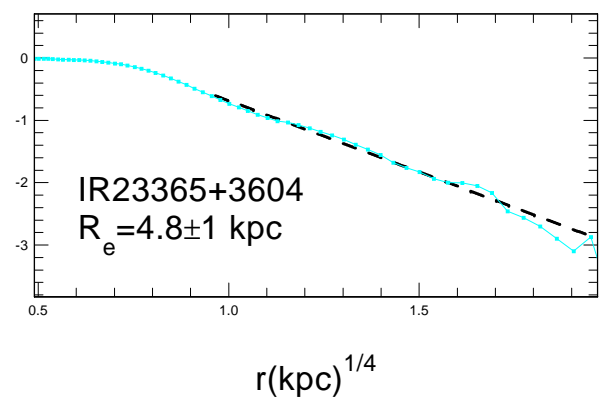

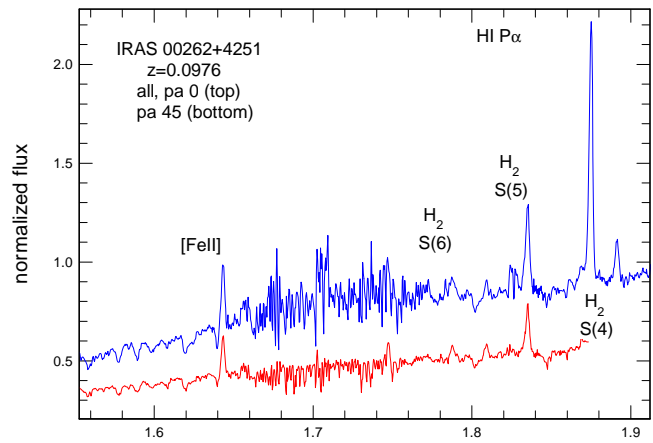
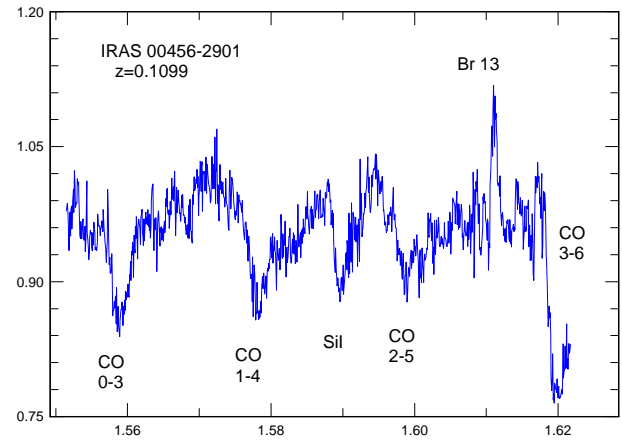
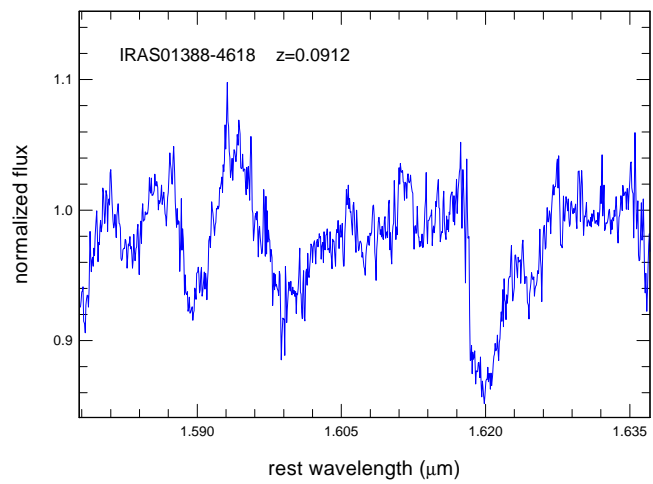
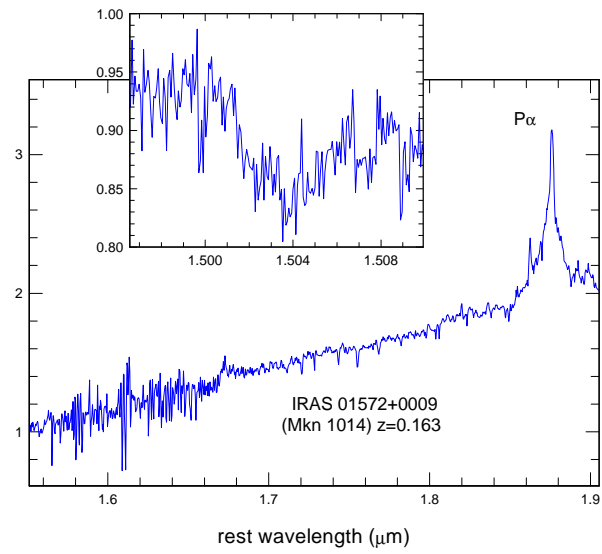

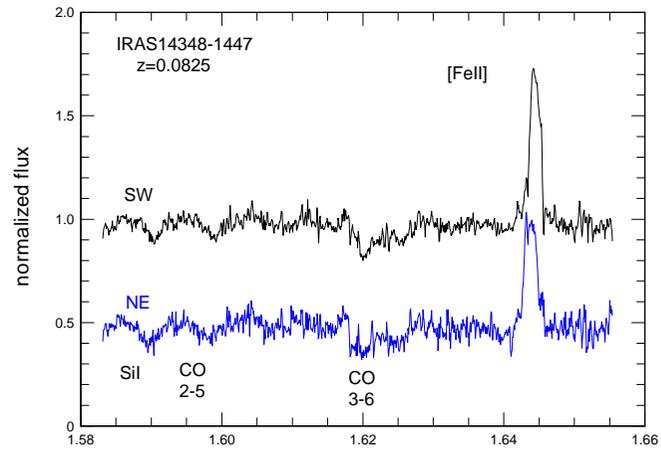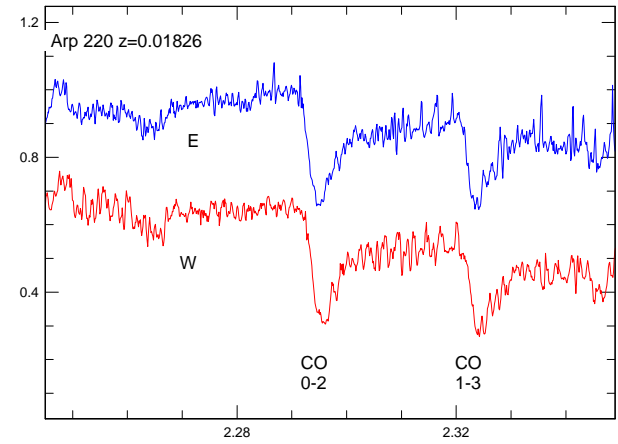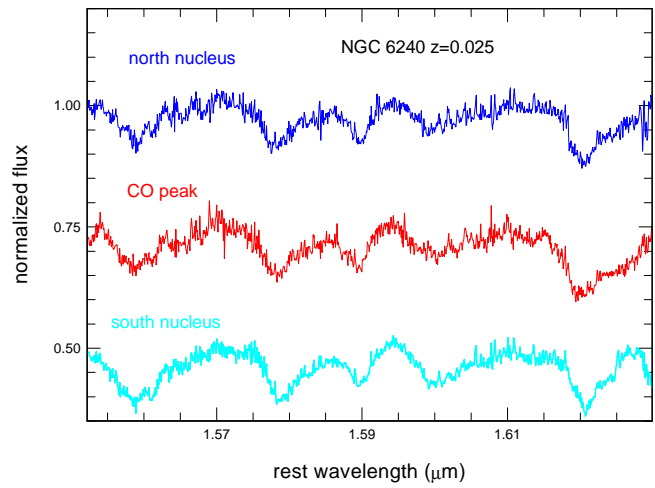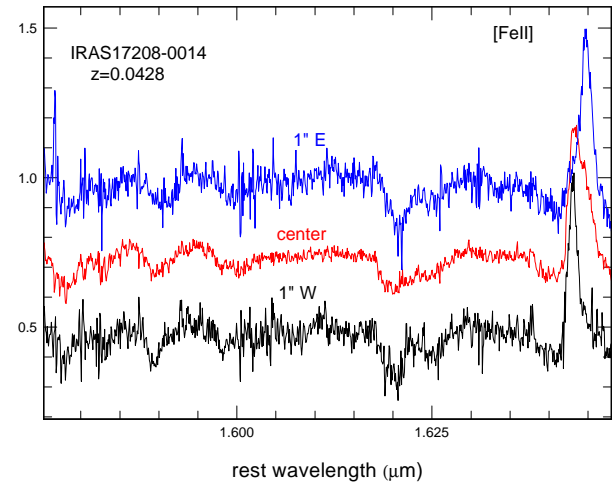

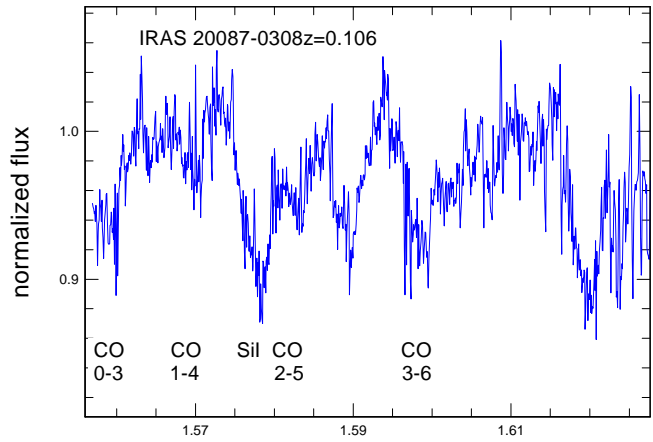
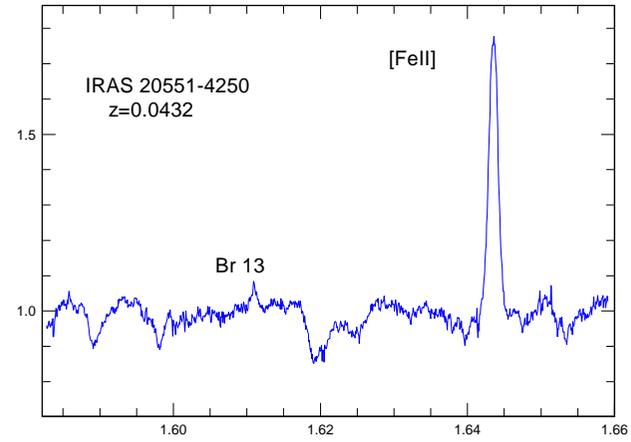
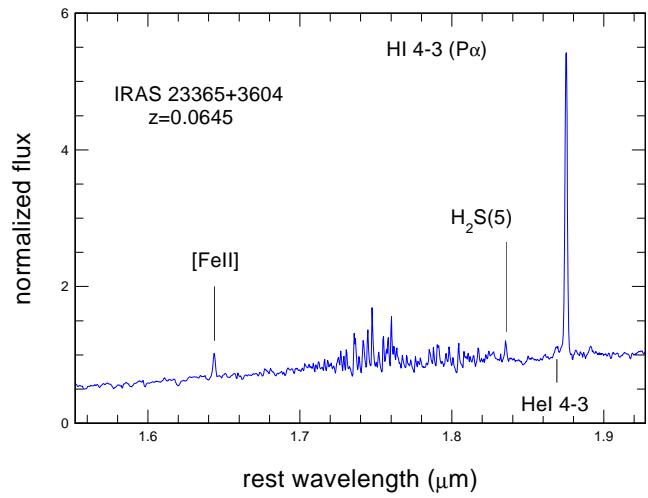
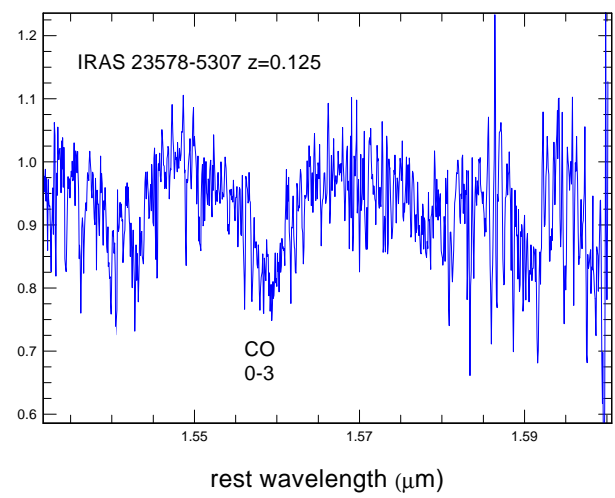

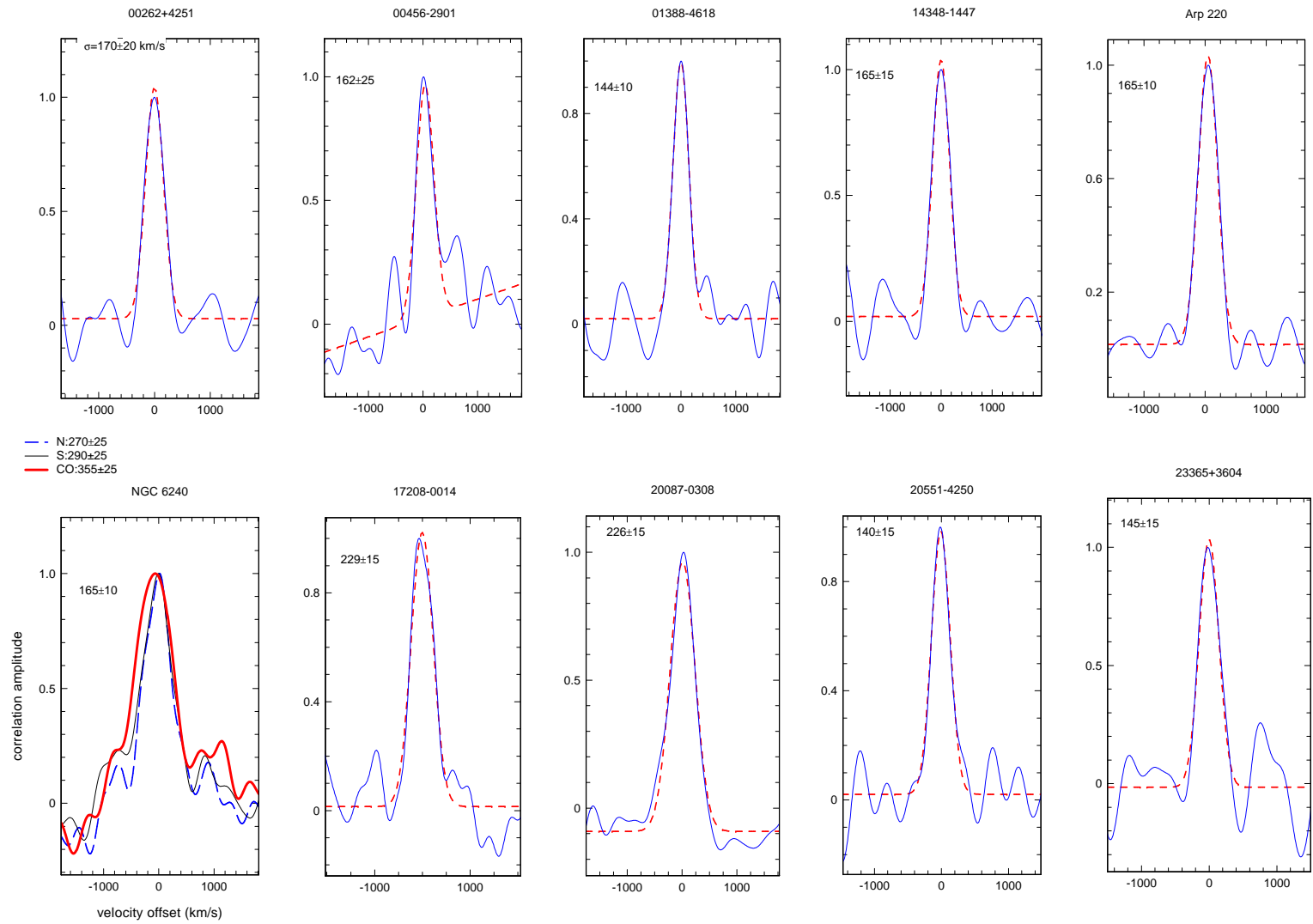

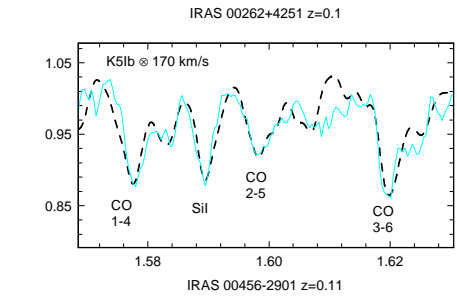
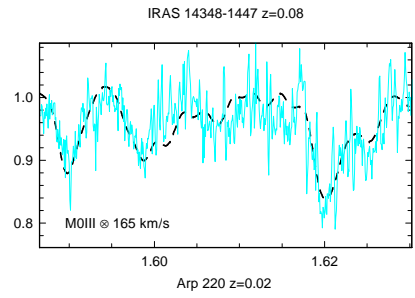
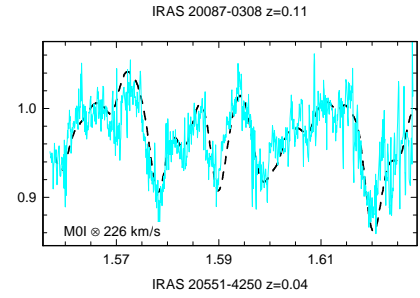
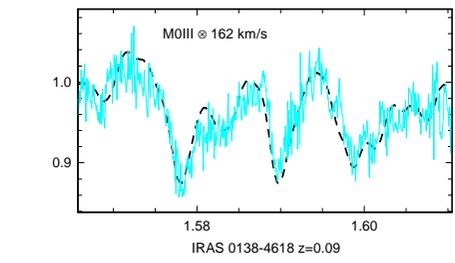
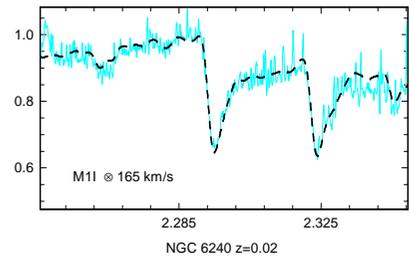
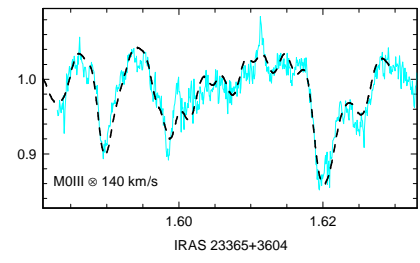
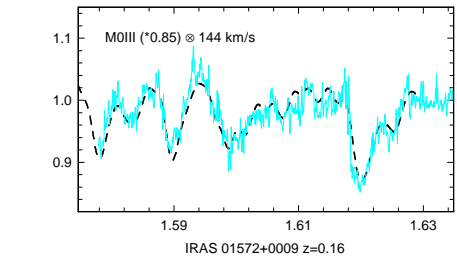
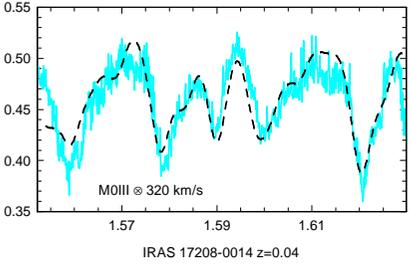
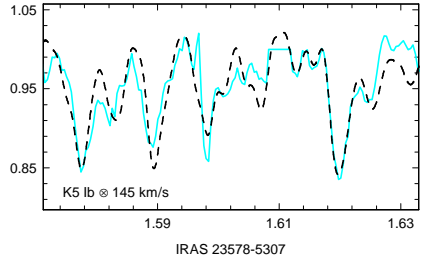
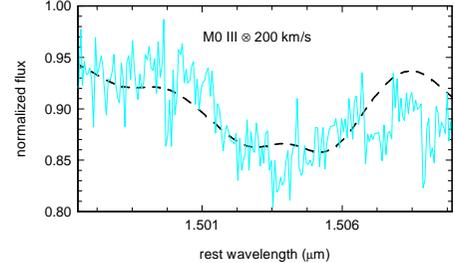
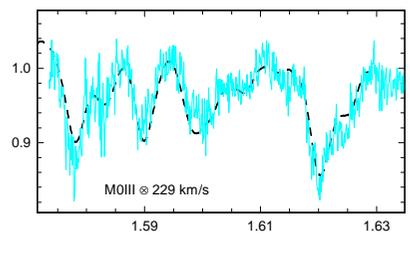
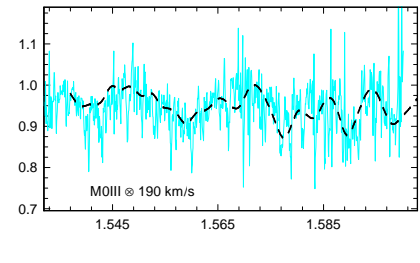

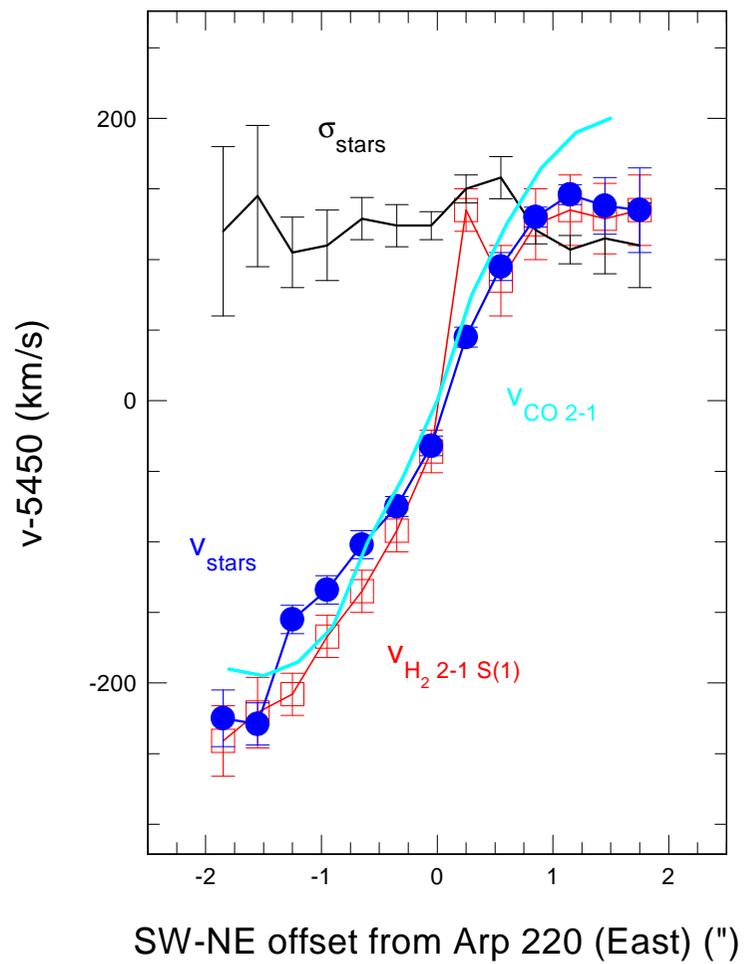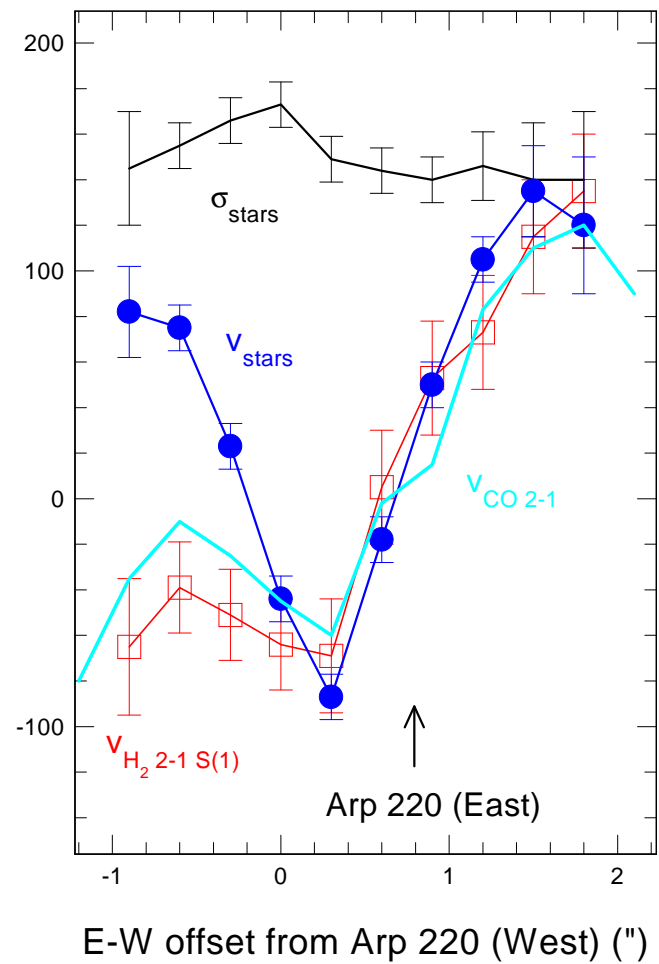

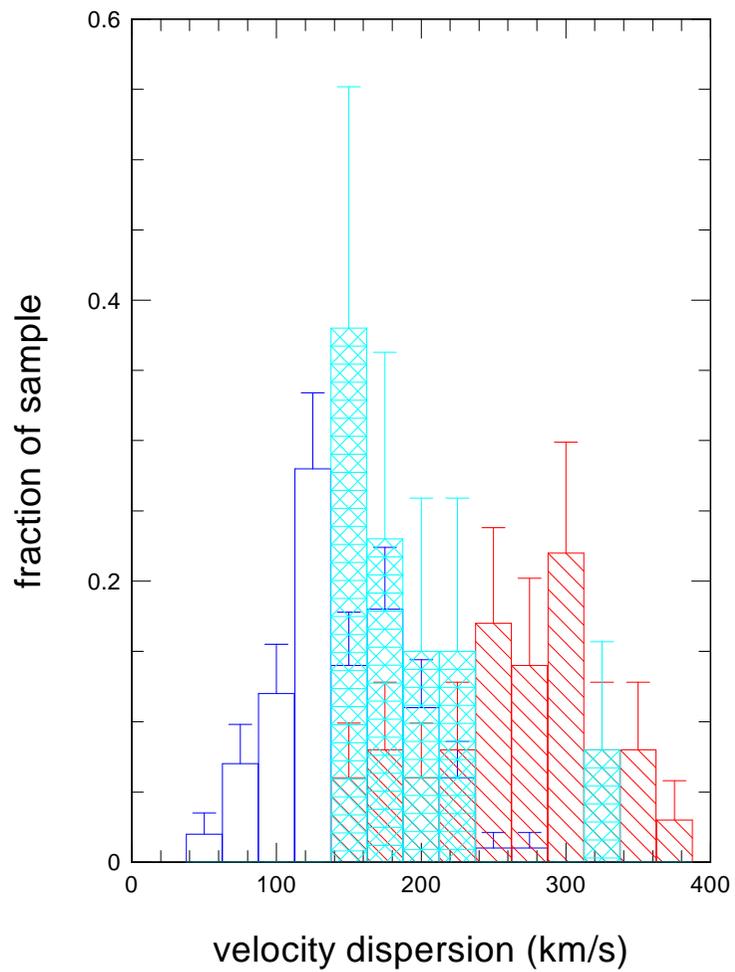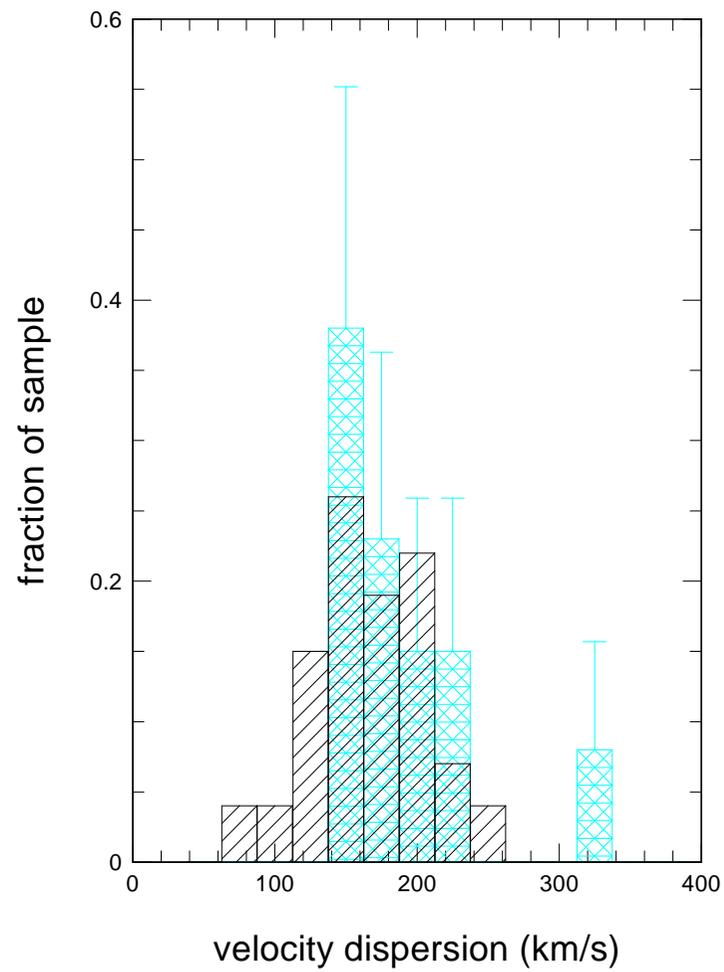

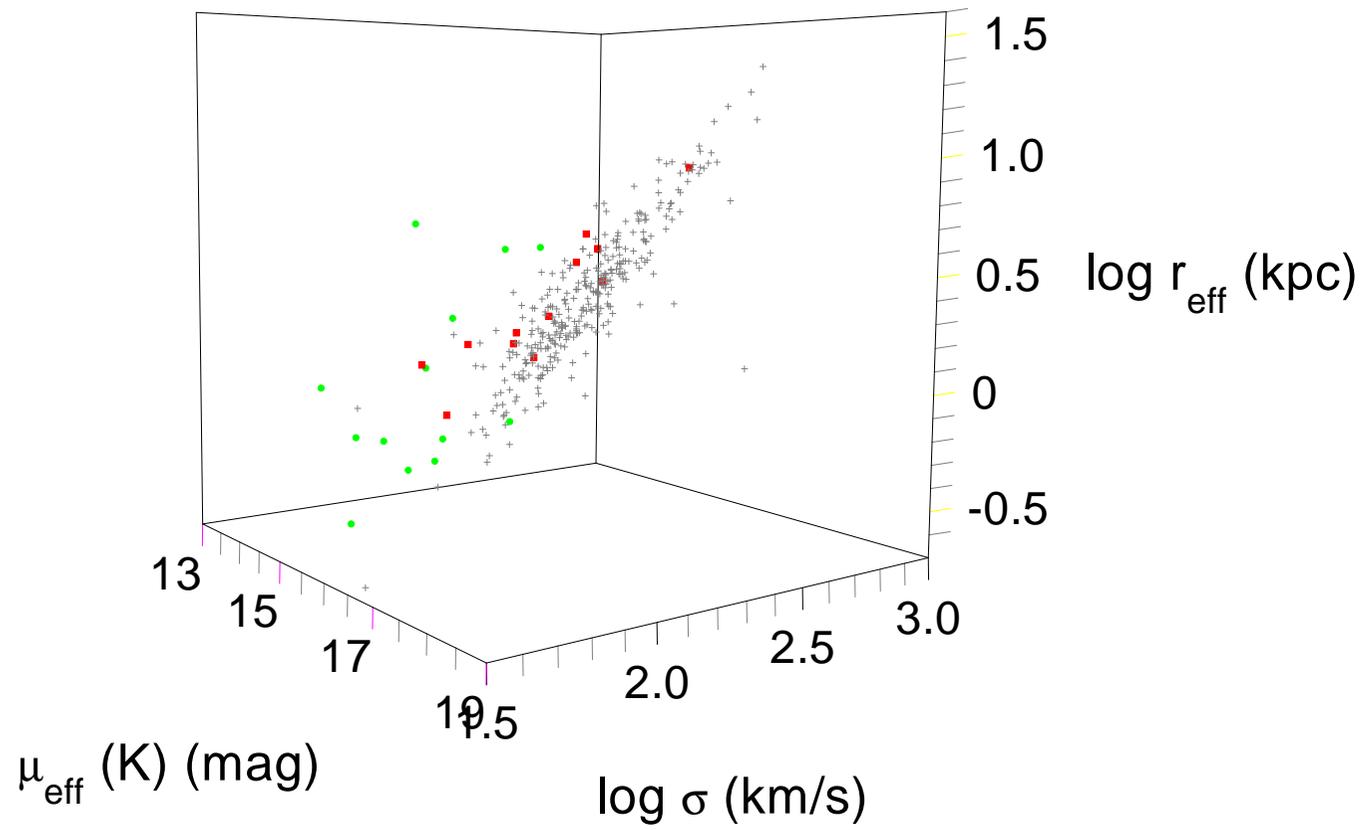

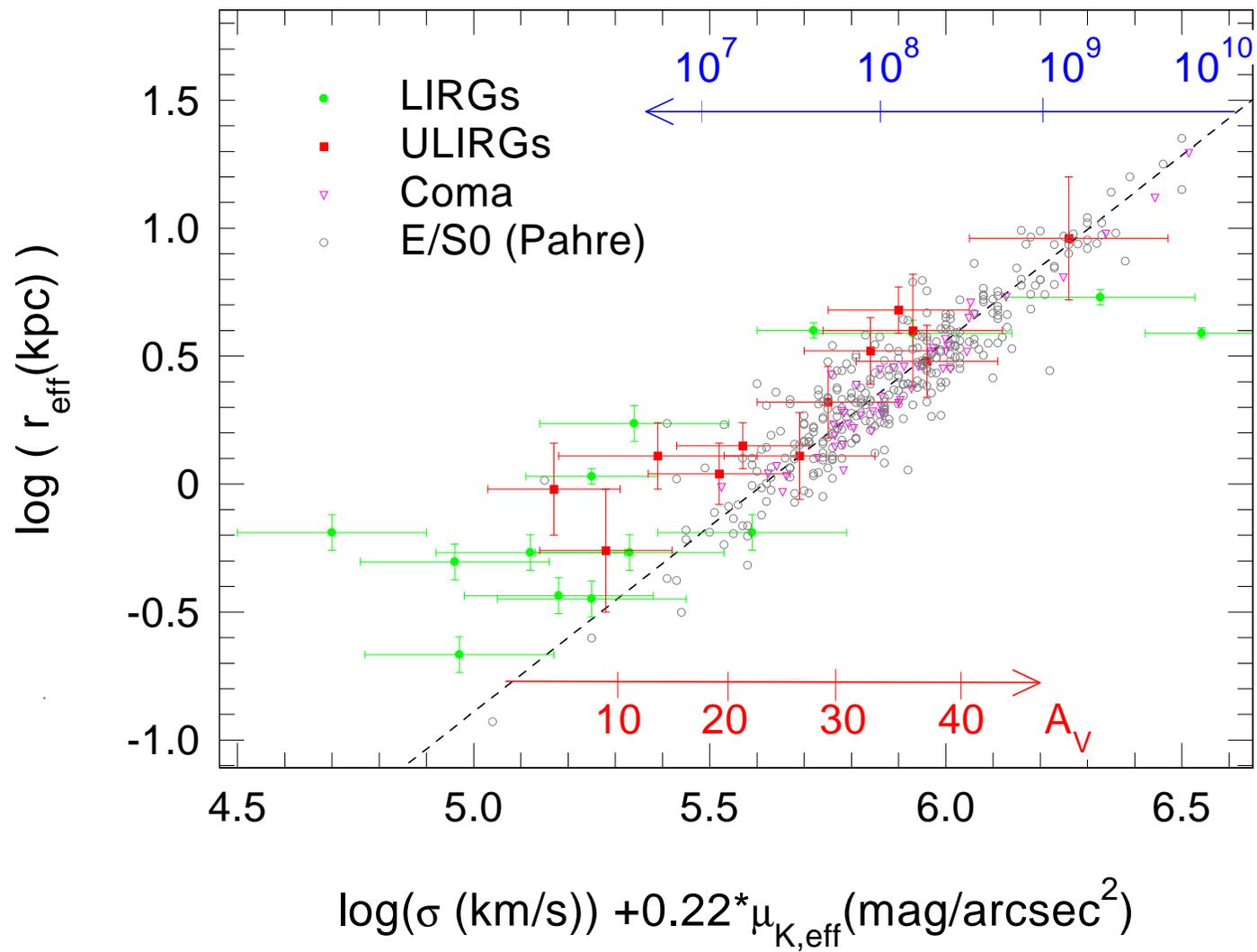

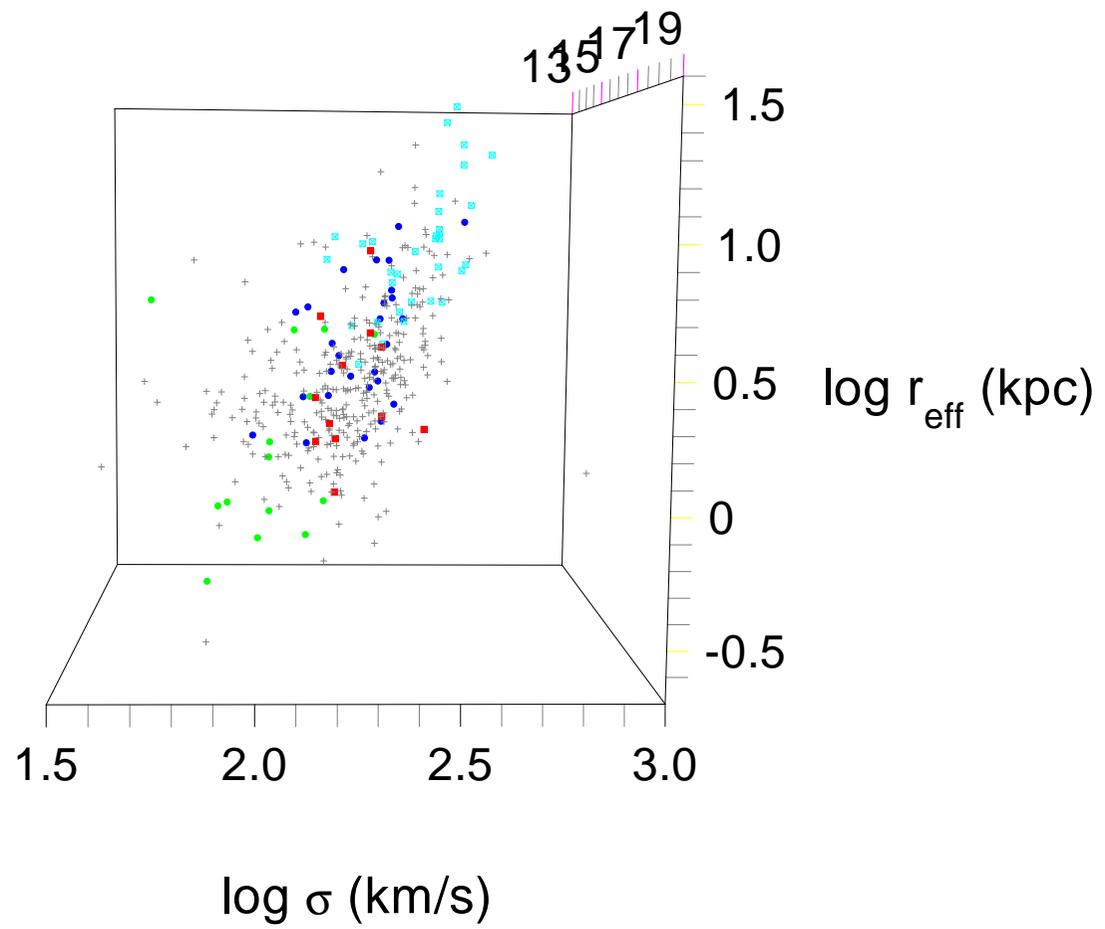

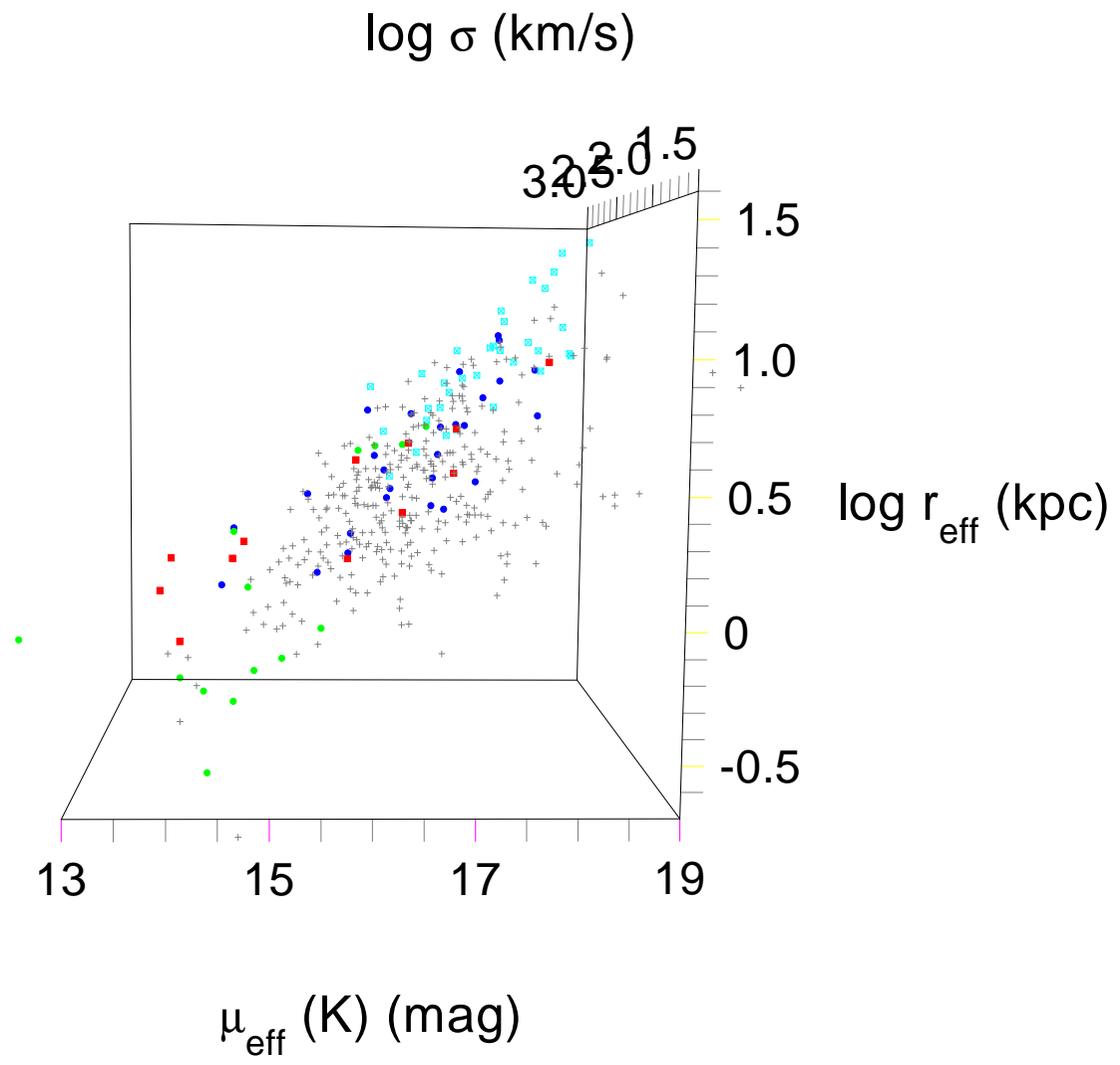

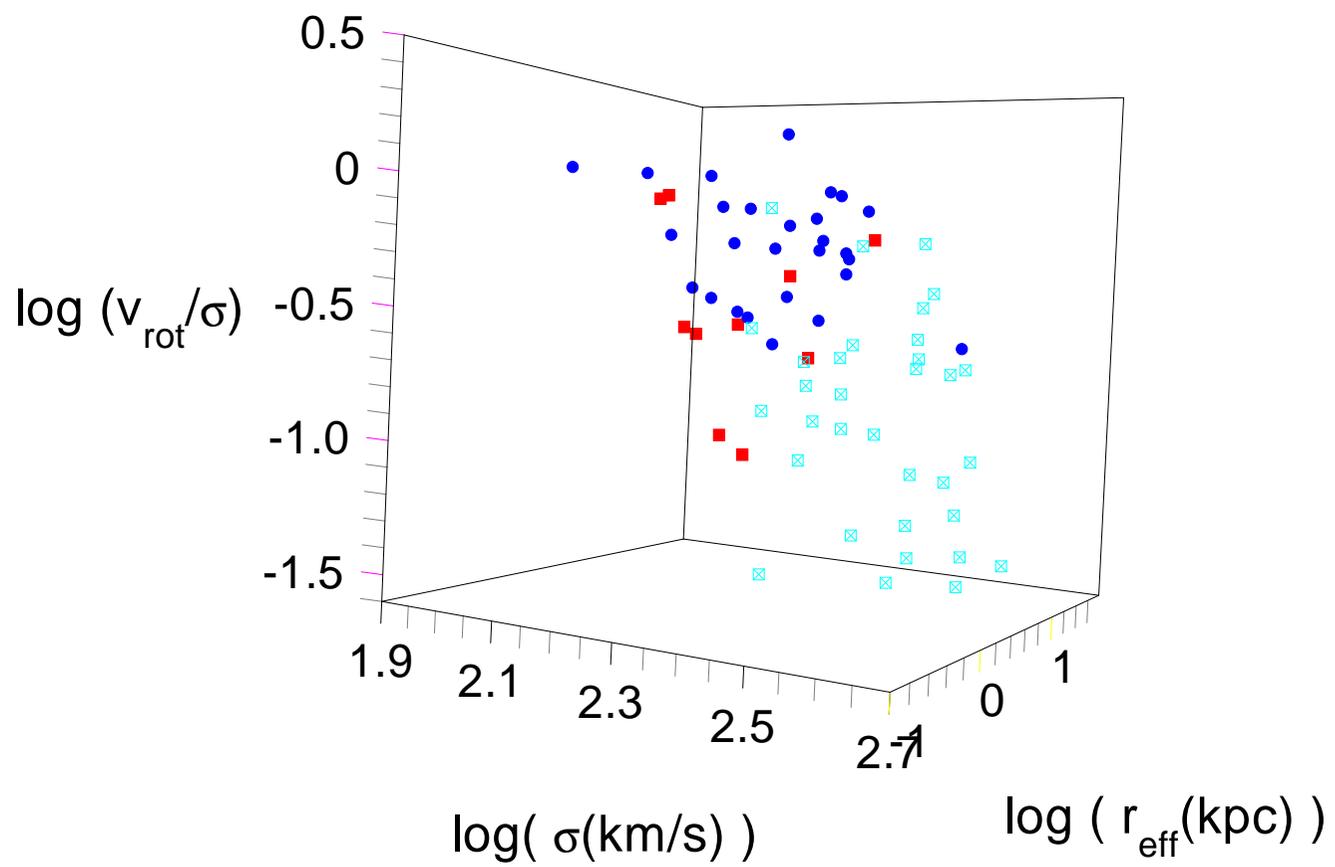

# Ultra-Luminous Infrared Mergers: Elliptical Galaxies in Formation?[1]

R.Genzel[2,3], L.J.Tacconi[2], D.Rigopoulou[2], D.Lutz[2] and M.Tecza[2]

[2]Max-Planck Institut für extraterrestrische Physik,
Garching FRG,
[3]Department of Physics, University of California, Berkeley USA

## ABSTRACT

We report high quality near-infrared spectroscopy of 12 ultra-luminous infrared galaxy mergers (ULIRGs). Our new VLT and Keck data provide ~0.5″ resolution, stellar and gas kinematics of these galaxies most of which are compact systems in the last merger stages.

We confirm that ULIRG mergers are 'ellipticals-in-formation'. Random motions dominate their stellar dynamics, but significant rotation is common. Gas and stellar dynamics are decoupled in most systems. ULIRGs fall on or near the fundamental plane of hot stellar systems, and especially on its less evolution sensitive, $r_{eff}$-$\sigma$ projection. The ULIRG velocity dispersion distribution, their location in the fundamental plane and their distribution of $v_{rot}\sin i/\sigma$ closely resemble those of intermediate mass (~$L_*$), elliptical galaxies with moderate rotation. As a group ULIRGs do not resemble giant ellipticals with large cores and little rotation. Our results are in good agreement with other recent studies indicating that disky ellipticals with compact cores or cusps can form through dissipative mergers of gas rich, disk galaxies while giant ellipticals with large cores have a different formation history.

---

[1] Based on observations at the European Southern Observatory, Chile (ESO Nos. 65.N-0266, 65.N-0289 ), and on observations at the W.M. Keck Observatory, which is operated as a scientific partnership among the California Institute of Technology, The University of California and the National Aeronautics and Space Administration. The Keck Observatory was made possible by the general financial support by the W.M. Keck Foundation.

*Subject headings: galaxies: elliptical and lenticular, cD – galaxies: formation – galaxies: kinematics and dynamics – infrared: galaxies*

# 1. INTRODUCTION

A significant new element in the exploration of star formation in the distant Universe has emerged with the COBE detection of an extragalactic far-IR/submillimeter background (Puget et al 1996). Its energy density is at least as large as that of the combined light of all galaxies emitting in the rest frame optical/ultra-violet bands. The far-IR/submillimeter background appears to be dominated by luminous ($>10^{11.5} L_\odot$) galaxies at $z \geq 1$ (e.g. Smail, Ivison & Blain 1997, Hughes et al. 1998, Barger et al 1998, Genzel and Cesarsky 2000). Very luminous and massive (Rigopoulou et al. 2001), dusty starbursts (star formation rates $10^2$ to $10^3$ $M_\odot yr^{-1}$) thus have probably been contributing significantly to the cosmic star formation rate at $z \geq 1$. These dusty starbursts may be large bulges/ellipticals in formation (e.g. Guiderdoni et al 1998, Blain et al 1999). (Ultra-) luminous infrared galaxies ((U)LIRGs[2]: Sanders et al 1988, Sanders & Mirabel 1996) may be the local analogues of the high-z population. Following the 'ellipticals through mergers' scenario of Toomre and Toomre (1972), Kormendy and Sanders (1992) have proposed that ULIRGs may evolve into ellipticals through merger induced, dissipative collapse. Invariably ULIRGs are advanced mergers of gas rich, disk galaxies (Sanders & Mirabel 1996). Their near-IR light distributions often fit an $r^{1/4}$-law (Wright et al. 1990, Stanford & Bushouse 1991, Doyon et al. 1994, Scoville et al. 2000). ULIRGs have large central molecular gas concentrations with densities comparable to stellar densities in ellipticals (Downes & Solomon 1998, Bryant & Scoville 1999, Sakamoto et al. 1999, Tacconi et al.

---

[2] In the definitions of Soifer, Houck & Neugebauer (1987) and Sanders & Mirabel (1996) 'ultra-luminous' infrared galaxies (ULIRGs) have infrared (10-1000μm) luminosities $>10^{12} L_\odot$. Galaxies with infrared luminosities $<10^{12}$ and $>10^{11} L_\odot$ are called 'luminous' (LIRGs). While these definitions are somewhat arbitrary, only the ULIRG-category is comprised almost exclusively of compact mergers of very gas rich and luminous galaxies.

1999). These gas concentrations (and stars forming from them), along with dense and massive, central bulges in the precursor galaxies (Hernquist, Spergel & Heyl 1993) may be the crucial ingredients for overcoming the fundamental phase space density constraints that would otherwise prevent formation of dense ellipticals from much lower density disk galaxies (Ostriker 1980). Observations and numerical simulations of mergers (e.g. Toomre & Toomre 1972, Barnes & Hernquist 1992) suggest an attractive and simple picture for the formation of massive ellipticals from mergers of gas rich disks, in the process initially resulting in very luminous starbursts and later evolving to QSOs (Sanders et al 1988).

During the last two decades ground-based optical and HST observations have established that nearby spiral bulges, S0s (lenticular galaxies) and ellipticals span a well defined, two-dimensional plane in velocity dispersion ($\log \sigma$)-effective radius ($\log r_{eff}$)-surface brightness ($\log \Sigma_{eff}$ or $\mu_{eff}$) space (e.g. Kormendy & Djorgovski 1989 and references therein). This 'fundamental plane' is a key characteristic of the near-homologous, near-constant M/L (~virial) equilibrium of hot stellar systems (e.g. Pahre, de Carvalho & Djorgovski 1998). There seem to be two distinct classes of hot galaxies, which populate different parts of the fundamental plane and, which also are distinct in the relative importance of stellar rotation and random motions (Faber et al. 1997). One class comprises spiral bulges and low to moderate mass and luminosity lenticular and elliptical galaxies with 'disky' isophotal shapes. The 'disky ellipticals' and lenticulars have power law surface brightness distributions to small spatial scales and relatively small effective radii. They exhibit significant rotational support ($v_{rot}/\sigma$~0.2-1.5). The second class consists of very luminous, massive (giant) ellipticals that have large effective radii, distinct large cores and 'boxy' isophotal shapes. These 'boxy ellipticals' are pressure supported with very little rotation ($v_{rot}/\sigma \ll 1$), mildly triaxial and anisotropic. A large fraction of boxy, giant ellipticals is found in the centers of galaxy clusters while disky ellipticals are found in all environments including the field (e.g. Quillen, Bower & Stritzinger 2000). Kormendy and Bender (1996) and Faber et al. (1997), among others, have proposed that disky ellipticals and boxy ellipticals have different formation histories. Disky ellipticals with small or no cores may be the result of relatively recent mergers of dissipative gas rich disk galaxies (Kormendy 1989, Nieto et al. 1991, Bender, Burstein & Faber 1992). Boxy ellipticals probably have formed a long time ago, perhaps through several merger hierarchies (e.g. Bender et al. 1992, Naab, Burkert & Hernquist 1999), or through the dynamical influence of a decaying, binary black hole (Faber et al. 1997).

If the *'elliptical-in-formation'*-scenario is correct, ULIRG mergers should lie on the fundamental plane for hot stellar systems. If boxy and disky ellipticals form through two entirely different formation paths (e.g. gas-poor, dissipation-less vs. gas-rich with dissipation) ULIRG mergers should resemble disky ellipticals but have different properties than boxy ellipticals. If boxy ellipticals are the result of their black holes gouging out large cores in the merger remnants (gas rich or not), ULIRG mergers may result in either boxy or in disky ellipticals. These different proposals can be tested by determining the structural properties of a sample of these systems and by placing these in $\sigma$-$r_{eff}$-$\mu_{eff}$ space. ULIRGs are dusty and their central few kpc regions are usually highly extincted ($A_V \geq 5$). Observations thus have to be made in the near-infrared band.

The fact that the near-infrared stellar surface brightness distributions of a significant fraction of the ULIRGs fit $r^{1/4}$-laws is consistent with the elliptical-in-formation-proposal. It is not sufficient or conclusive, however. ULIRGs as a rule have prominent tidal tails on large scales and bright star forming regions, or dusty active galactic nuclei at their centers. Their combination could qualitatively mimic an $r^{1/4}$-distribution. Differential dust extinction may distort their surface brightness distributions (e.g. Scoville et al. 2000).

Observations of the *stellar dynamics* are the key for unraveling the structural properties of ULIRG mergers. So far infrared spectroscopy on 4m-class telescopes has been limited to a few very bright (U)LIRGs. Aperture spectroscopy of Arp220 and Mrk273 indicate moderately high velocity dispersions ($\sigma \sim 150$ km/s, Doyon et al. 1994, James et al. 1999). The luminous merger NGC6240 has a velocity dispersion akin to the most massive giant elliptical galaxies (300-350 km/s in a ~2" aperture encompassing both nuclei: Doyon et al. 1994, Lester and Gaffney 1994). These findings, and in particular the result for NGC6240, are consistent with the elliptical-in-formation-scenario.

While ULIRGs and mergers as a class are transitory objects undergoing rapid evolution, late stage mergers probably provide a fair sample of the dynamical properties of the systems they finally evolve into. Recent numerical simulations of galaxy mergers suggest that violent relaxation is very effective. By the time the two nuclei merge within less than about 1 kpc and then rapidly coalesce, the systems have basically reached their equilibrium values of rotation, dispersion and even higher order kinematic moments (skewness and kurtosis) on spatial scales of the half mass radius or greater (Mihos 1999, Bendo & Barnes 2000).

In the present paper we report, for the first time, high quality stellar (and gas), near-IR spectroscopy of a substantial sample of ULIRGs (12 galaxies). Our program has become possible with the advent of high quality infrared spectrometers on the VLT and Keck telescopes, in combination with sub-arcsecond seeing.

## 2. OBSERVATIONS AND DATA REDUCTION

The data were taken with the ANTU telescope of ESO's VLT on Cerro Paranal, Chile and with the Keck 2 telescope on Mauna Kea, Hawaii. At the VLT the data were taken with the facility near-infrared camera and spectrometer ISAAC (Moorwood et al. 1998) in two observing runs in spring and summer 2000. At the Keck 2 telescope the data were taken with the facility spectrometer NIRSPEC (McLean et al. 1998) in summer 2000. NIRSPEC was in low resolution mode and had $\lambda/\Delta\lambda=R(FWHM)=2200$, a slit width of 0.58" and 0.193" per pixel along the slit. ISAAC was in mid-resolution mode (R(FWHM)=5200), with a slit width of 0.6" and 0.147" per pixel along the slit.

We selected our program galaxies fitting the ULIRG criteria ($L_{IR} \geq 10^{12}$ $L_\odot$) from the BGS (Sanders et al. 1988) and 1Jy (Kim and Sanders 1998) catalogues, and in a few cases from the Faint Source Survey (Moshir et al. 1992) catalogue. We culled from these IR-luminosity selected catalogues those sources that fitted our RA/Dec-boundary conditions and that were compact, single nuclei systems on near-IR images (but with definite signatures for a recent merger, such as tidal tails). Such systems are very likely late stage, merger remnants (see Introduction). We picked only those sources with redshifts ≤0.16 that could be observed in the H-band (see below). This set of selections reduced the number of available sources from about 70 fitting the RA/Dec-requirements to about 15. We were able to get H-band spectra of 9 of these. In addition (and partially as a control group) we observed three compact double nuclei ULIRGs: Arp220, IRAS14348-1447 and NGC6240. These ULIRG mergers are all located in the field or in small groups; none are situated in dense clusters. NGC6240 (projected nuclear separation 800 pc) and Arp 220 (separation 350 pc) are late stage mergers as well, although their nuclei obviously have not coalesced yet. With a projected nuclear separation of 5.5 kpc IRAS 14348-1447 is probably in a somewhat earlier merging stage. NGC6240 does not quite make it formally into the ULIRG category (its infrared luminosity is log $L_{IR}$=11.8) but it has all other characteristics of a very luminous merger, and we took the present H-band spectroscopy for comparison with our previously published K-band spectroscopy with the MPE 3D-spectrometer (Tecza et al. 2000). The total program consisted of 5 nights of observing time of which 4.5 were useable and essentially all of that time was in very good to excellent seeing conditions (0.35 to 0.8" FWHM in the H/K-band). Table 1 gives the source list. All distances in this paper are for a $H_o$=70 km/s/Mpc, $\Omega$=0.3 cosmology.

There are several ways to probe stellar absorption features in the near-IR. The $\Delta v=2$ overtone, ro-vibration bands of CO at 2.29-2.4μm have the largest equivalent widths but become inaccessible from the ground at modest redshifts ($z \geq 0.04$). The 2-3 times weaker $\Delta v=3$ overtone bands (plus features due to SiI, OH etc.) in the 1.56-1.64μm region can be observed in the H-band out to about z~0.15, and then again in the K-band for $z \geq 0.25$ (Figure 1). In the J-band there are various fainter features that can be observed at low redshift, as well as the strong 8600 Å Ca-triplet for $z \geq 0.4$. Since most of the ULIRGs are at redshift $\leq 0.2$, we chose to observe the rest-wavelength H-band features for all galaxies except Arp 220 (z=0.018), in which we observed the $\Delta v=2$ K-band CO bands. To obtain access to the deepest H-band features (CO, SiI) we chose from the catalogs those galaxies with redshifts <0.15 (with the exception of the bright QSO-like ULIRG Mkn 1014 (IRAS 01572+0009, z=0.16)).

For the data reduction we used standard IRAF routines and scripts. We took the basic spectroscopic frames (on-chip integration time 240-600 seconds) in nod-along-the-slit-pairs. For sky background removal (including OH-lines) we subtracted the two frames of a pair and then flat-fielded the result, also creating a negative (sign-reverted) copy of each subtracted/flat-fielded frame. We created a 'star' frame with two bright stars in different positions along the slit for determining the spatial correction matrix, and either a dark-subtracted sky-frame (with bright OH sky lines) or a Neon-Argon arc-lamp frame for determining the wavelength correction matrix. Following removal of hot/bad pixels in the sky, star and flat field frames, we then rectified in spatial and wavelength coordinates and simultaneously wavelength calibrated all frames. Both NIRSPEC and ISAAC spectra exhibit significant curvature in both spatial and wavelength coordinates. Next we carried out a first removal of hot/bad pixels in the source frames, determined the flux weighted peak position for the (positive) source in each frame, and shifted all frames to a common centroid. After another iteration of bad pixel/cosmic ray removal we then arrived at the final result for a given slit position by dividing the sum frame by the extraction of an atmospheric calibration star (reduced in the same manner as above). For this purpose we selected either relatively featureless early type (late O or early B-stars) stars or G2 V stars near the elevation of the program sources. In the case of the early type stars we removed by Gaussian/Lorentzian fitting the few intrinsic features (mostly Brackett series hydrogen lines). In the case of the G2V stars we divided by an appropriately Doppler-shifted, high-resolution solar spectrum smoothed to our instrumental resolution.

We also observed and reduced a sample of spectral calibrator stars with setups identical to those used for the program galaxies. For these stars we selected K5-M1 giants and supergiants. These stars are known to be good templates for typical luminous, star-forming galaxy spectra where the contribution from dwarfs can be neglected (e.g. Origlia, Moorwood & Oliva 1993, Oliva et al. 1995). There are no strong differences between the H-band properties of late type giants and supergiants, and our results are insensitive to possible differences in stellar populations (Figure 2). In the end, we chose the M0 III giant HD25472 as spectral template for the VLT data, and the K5 Ib supergiant HD200576 as template for the Keck observations (Figures 1, 2). In the case of Arp220 we picked the M1 Iab supergiant HD99817, which is an excellent match to its spectral properties.

For most sources we observed along two different position angles placed either along the isophotal major and minor axes, or determined from prior knowledge of the dynamical properties (e.g. Arp 220, NGC 6240 and IRAS 17208-0014). In this way we could test for the importance of rotation in the merger stellar dynamics. We extracted spectra in selected apertures from the final data set. These spectra were then normalized and continuum subtracted for further quantitative comparison with the spectral template stars. Wherever necessary we also removed residuals (=shot noise) of strong OH sky emission lines by replacing the flux at the OH wavelength with the average from either side of it. For velocity determinations we used a cross-correlation of the logarithmically re-sampled program galaxy and stellar template spectra. For determinations of velocity dispersions and line-of-sight-velocity-distributions (LOSVDs) we used a Fourier-quotient technique. We divided the galaxy-stellar template cross-correlation in Fourier space by an auto-correlation of the stellar template, using a Wiener filter to suppress the high-frequency noise (c.f. Bender 1990). For this purpose we used a code written by one of us (R.G.) and further improved by Anders (1999). Wherever possible, we used the entire wavelength range with reasonably deep features for the correlation (rest frame 1.57-1.63μm in the case of the H-band, 2.25-2.35μm in the case of the K-band). Gaussian fits to the final LOSVDs give the dispersions shown in Figures 5 and 6 and listed in Table 2. Their uncertainties are dominated by systematic effects, such as template mismatch, incompletely removed atmospheric and stellar features, OH shot noise and emission lines. These systematic errors typically add up to ±15 to ±25 km/s. We derived rotation velocities from cross-correlations with the template stars to spectra extracted ±1" to ±2" from the nuclei, equivalent to ±0.5-1x $r_{eff}$ (Table 2).

For the derivation of effective radii of the near-infrared surface brightness distributions we took H-band images of 6 of our program galaxies at the Keck and VLT. Using the 'Ellipse' routine in STSDAS we then fitted elliptical isophotes to the resulting sky-subtracted and flat-fielded images. To obtain effective radii we then fitted $r^{1/4}$-(de Vaucouleurs) profiles, typically between $r^{1/4}$=0.55 to 1.6 ($kpc^{1/4}$). IRAS 01572+0009 has a bright active galactic nucleus (AGN) that we excluded from the fit. In addition we analyzed in the same manner the K-band, NICMOS surface brightness distributions of 5 of our sources from the work of Scoville et al. (2000), which Nick Scoville kindly made available to us. For 3 sources we analyzed the ground based, K-band surface brightness distributions from the analysis of the 1 Jy sample of Veilleux, Kim & Sanders (2001), which were kindly made available to us by Sylvain Veilleux and Dave Sanders. For IRAS 23365+3604 we analyzed the H-band image of Davies et al. (2001). Our new imaging results are displayed in Figure 3.

# 3. RESULTS

*3.1 Structural and dynamical properties of the ULIRG sample*

The basic results of our study are contained in Figures 3-8 and summarized in Table 2. Figure 3 shows the H-band surface brightness distributions as a function of $r_{major}^{1/4}$ for six of the galaxies for which there was no such information in the literature, along with the derived effective radii of the best fit, de Vaucouleurs law. Figures 4 through 6 display our final spectra, deduced stellar LOSVDs and comparisons of the galaxy spectra with stellar templates convolved with the best fit Gaussian fits to the LOSVDs. Figure 7 shows a comparison of gas and stellar rotation velocities and stellar velocity dispersions in Arp 220. Figures 8a and 8b display position velocity diagrams for emission lines ([FeII] 1.64μm, 1.875μm HI Pα, 2.24 v=2-1 S(1) $H_2$) for 7 of our program galaxies. Table 2 summarizes the derived stellar velocity dispersions, rotation velocities for stars and gas and effective radii. Appendix A contains comments on individual sources. All quantities are for $H_0$=70 km/s/Mpc and $\Omega_0$=0.3.

Most of the merger remnants have compact near-infrared light distributions, with effective radii of a few kpc or less. The effective radii listed in Table 2 are quite uncertain (±30 to ±50%), as a result of differential extinction, incomplete relaxation and population effects. The images show significant deviations from ellipsoidal shapes in many sources (including the tidal tails: Duc et al. 1997, Rigopoulou et al. 2000, Scoville et al. 2000, Surace et al. 2000, Veilleux, Kim and Sanders 2001). Deep K-band CO absorption features in several sources indicate that the stellar light is dominated by relatively young stars (Armus et al. 1995, Goldader et al. 1995, Tecza et al. 2000). The greatest uncertainty is the effect of extinction, which is substantial in most of our ULIRGs even at 2µm. Equivalent screen, K-band extinctions range between 0.4 and >1 mag, as determined from near-infrared colors or infrared line ratios (Scoville et al. 2000, Genzel et al. 1998). Scoville et al. (2000) also find that 1.1 to 2.2µm colors redden towards the nuclear regions of most their ULIRGs studied with NICMOS. The most obvious explanation of this finding is increased extinction there relative to the outer parts. Other possibilities may be nuclear hot dust emission (e.g. from a central AGN, as in IRAS01572+0009), or a circum-nuclear concentration of young, cool supergiants. We believe that for most of our sources differential dust extinction is the most likely explanation for the red nuclear colors. For determining effective radii we therefore used K-band data if available, but also excluded the nuclear regions (<400 pc) in fitting the effective radii (see Fig.3). J- or H-band effective radii tend to be significantly larger than those determined from K-band data, typically by a factor of 2 in the case of the NICMOS data of Scoville et al. (2000). We have also corrected the K-band surface brightness distributions of Scoville et al. (2000) with extinctions derived from the 1.1 to 2.2µm brightness ratios. However, we found the additional correction to the effective radii to be relatively small, with the exception of Arp 220 (where the extinction corrected $r_{eff}$ is 0.4 kpc as compared to 0.56 kpc derived from the uncorrected K-band data).

For the comparison of the fundamental plane properties (section 3.4), we have also compiled in Table 2 rough K-band magnitudes within r=$r_{eff}$, $\mu_{eff}(K)$, as determined from the surface brightness distributions. These magnitudes are necessarily quite uncertain, since in about half the cases we had to use the H-band surface brightness distributions, with flux calibrations on source integrated, K-band data from the literature. These magnitudes are also subject to the full effects of extinction and time evolution of the underlying populations. The K-surface brightness per solar mass brightens by about 2.2 mag from an old (t>$10^{10}$ yrs) population to one in the AGB phase (t~a few $10^8$ yrs) and then, another 1.7 mag to one dominated by young, red supergiants (t~$3 \times 10^7$ yrs, Bruzual & Charlot 1993).

## *3.2 Lower luminosity, infrared bright mergers*

Shier and Fischer (1998) and James et al. (1999) have studied a sample of 13 infrared bright mergers of luminosity $10^{11}$ to <$10^{12}$ $L_\odot$ (LIRGs). For comparison to our ULIRG sample we list in Table 3 their corresponding structural and dynamic properties.

## *3.3 ULIRG mergers are relaxed with significant rotation*

On average the stellar dynamics in our ULIRG sample is relaxed but rotation does play a significant role. In 6 of 7 single nucleus mergers the random motions dominate ($v_{rot}\sin i/\sigma$<1.7), although rotation is detected at some level in 5 of them. In 3 of 10 single and double nuclei galaxies rotation is equally or more dominant than random motions: 2 of them are the double nuclei systems Arp 220 and NGC6240. In 8 galaxies our near-IR emission line data (Figures 8a and b) and/or CO millimeter interferometry (Tacconi et al. 1999, Tecza et al. 2000 for NGC 6240, Downes and Solomon 1998 and Sakamoto et al. 1999 for Arp 220, and Downes and Solomon 1998 for IRAS 17208-0014 and IRAS23365+3604) give good information on gas motions. In 6 of these 8 galaxies the gas exhibits substantial rotation, often on spatial scales larger than the stellar emission. The emission line position-velocity diagrams in Figures 8a and b are consistent with the interpretation that the gas originates in kpc scale, rotating rings (Downes and Solomon 1998). In 4 of these 8 systems (and 4 of the 5 single nuclei systems) the gas rotational velocities are substantially greater than that of the stars. These findings suggest that gas and stellar motions are often decoupled, as exemplified in the complex spatial and kinematic structure of NGC 6240 (Tacconi et al. 1999, Tecza et al. 2000). Such a decoupling is expected if the stars have undergone strong dynamical heating due to violent relaxation while the gas dissipates to a cold state on a dynamical time scale. We conclude that late stage ULIRG mergers do resemble early type galaxies in their dynamic properties. It is interesting to note that fairly massive gas disks exist in these late stage merger systems. Naab and Burkert (2001b) have proposed that remnant gas disks surviving the merger process may explain the higher order kinematic moments of rotating ellipticals.

Since the program galaxies are fairly compact objects one might wonder, however, to what extent beam smearing effects (due to finite slit size and seeing) could have created the impression of random motions to be dominant while in reality the galaxies are rotating rapidly, or may even consist of two nuclei. Simulations with model galaxies show that beam-smearing effects indeed can lower the observed rotational amplitude significantly. An extreme example is to take the very compact Arp 220 system and move it to z~0.1, a distance characteristic of a number of the other systems. In that case and at 0.5" seeing and 0.6" slit size the source would be an unresolved single object and the counter-rotation pattern in the east-west slit of Figures 7 and 8b would no longer be recognizable. The rotation of the eastern nucleus along p.a. $52^o$, however, would still be observable with sufficient signal to noise, albeit at about 2 to 3 times lower amplitude. A more realistic simulation of beam smearing is an inclined rotating disk that fits the observed FHWM of the sources (in the spectroscopic data typically 1-2" FWHM along a slit). In that case the reduction of the observed rotation velocity due to beam smearing is between 10 and 30%, a significant but not dominant effect. We conclude that beam smearing cannot have caused our program galaxies to look like relaxed early type galaxies. However, the rotation velocities listed in Table 2 are *lower limits*, due to the combination of limited signal to noise ratios, beam smearing and inclination effects. In particular, we will underestimate the rotational component if the rotation curves still rise beyond the extraction radii of the rotation velocities, which ranged between 1 and 2", or 0.5-1x $r_{eff}$. Rising rotation curves beyond $r_{eff}$ are found in recent merger simulations (Mihos 1999, Bendo and Barnes 2000).

*3.4 Distribution of velocity dispersions*

Figure 9 shows a histogram plot of the measured *total* velocity dispersions of our ULIRG sample, plus the ULIRG Mkn 273 from the work of James et al. (1999). For the double nuclei IRAS14348-1447, Arp 220 and NGC 6240 we use the source integrated velocity dispersions (see notes to Table 2). With the exception of NGC 6240 (and perhaps Arp 220) none of the sources shows a significant difference between total/half light and central velocity dispersions. With the exception of NGC 6240 (see also Tecza et al. 2000) we also do not find any significant (>10-20% of peak) deviations of the LOSVDs from a Gaussian profile. For comparison with our ULIRG sample we show similar distributions for a sample of 44 giant ($M_B$<-20.5) elliptical galaxies, and 60 intermediate mass ellipticals (-18.5> $M_B$>-20.5) and lenticular galaxies (S0s) from Faber et al. (1997) and from Bender et al. (1992). We also show the distribution of velocity dispersions of 92 spiral bulges from Whitmore, McElroy & Tonry (1985). We list the median and mean velocity dispersions of these samples in Table 4.

*The mean/median velocity dispersion of our ULIRG sample is similar to that of an L\** *elliptical ($\sigma$(L\*)~180 km/s, $M_B$(L\*)~-20.4; Binggeli, Sandage and Tammann 1988; Davies et al. 1983; all adjusted to $H_o$=70).* The distribution of ULIRG velocity dispersions is very close to (and statistically indistinguishable from) those of intermediate mass ellipticals and lenticular galaxies. *As a group ULIRG mergers clearly have lower velocity dispersions than giant ellipticals.* The velocity dispersion distribution of ULIRGs also closely matches that of rotating/compact core/disky isophote ellipticals with intermediate masses, and is very different from the (more massive) slowly rotating/large core/boxy isophote ellipticals. We cannot formally exclude, however, that ULIRG mergers do sample the entire range of velocity dispersions of ellipticals above a minimum velocity dispersion of 140 km/s~0.77 $\sigma_*$ (~0.45 $L_*$), the lowest dispersion we see in our sample. This is because giant ellipticals are much rarer than $L_*$ ellipticals. Taking a Schechter luminosity function $\Phi(L)dL \sim (L/L_*)^\alpha \exp(-L/L_*)dL$ and a Faber-Jackson relationship $L/L_* = (\sigma/\sigma_*)^n$ with $\alpha=-1$ (Binggeli et al. 1988) and n=3.2 (Davies et al. 1983), less than 10% of a complete sample of ellipticals above $0.77 \times \sigma_* = 140$ km/s would have velocity dispersions above $\sigma_* = 180$ km/s. Thus in a sample of thirteen galaxies one would expect to see about one giant elliptical, consistent with our limited statistics.

ULIRGs have higher velocity dispersions than spiral bulges (Fig.9). That difference is, however, marginally significant (~2$\sigma$) if only the more massive early type (Sa, Sab and Sb) spirals are considered. This is not surprising since at least in the 3 double nuclei systems we have studied, the velocity dispersions of the two individual nuclei are fairly large and similar to the dispersions of the overall ULIRG sample. The large dispersions and estimated masses of these nuclei indicate that the progenitor galaxies were very likely intermediate to early type spiral galaxies with massive central bulges themselves.

We have applied a Student t-test to the velocity dispersion distributions to explore the hypothesis that ULIRGs and the other types of galaxies in Table 4 and Figure 9 are drawn from the same distribution (same mean and sample variance). The probability is high (20%) that ULIRGs, intermediate mass ellipticals, lenticulars and intermediate mass disky ellipticals are drawn from the same parent distribution. In contrast that probability is vanishingly small ($6 \times 10^{-5}$) for ULIRGs and giant/boxy ellipticals. The probability for all spiral bulges (Sa-c) and ULIRGs to sample the same distribution is also small (0.6%), while it is somewhat greater for the Sa-b bulges and ULIRGs (1.3%). ULIRGs thus have velocity dispersions that are matched only by $L_*$ early type galaxies and the most massive spiral bulges. Our conclusions do not change if the four double nuclei systems (Mkn 273, IRAS 14348-0014, Arp 220 and NGC 6240) are removed from the sample.

The distribution of the ULIRG velocity dispersions is similar to the merger remnants of Lake and Dressler (1986) drawn from the Arp and Arp-Madore atlases (Table 4), indicating that these objects may be more evolved versions of the same population. The ULIRGs have significantly higher velocity dispersions than the LIRG mergers of Shier and Fischer (1998) and James et al. (1999), suggesting that the LIRGs are the lower-mass continuation/tail of the ULIRG population.

*3.5 ULIRG mergers are on/near the fundamental plane*

In the following we will place the ULIRG (and LIRG) mergers in the space spanned by $\sigma$, $r_{eff}$ and $\mu_{eff}$. In this space dynamically hot galaxies (ellipticals, bulges and lenticulars) are found in a fairly well defined plane, as a result of the virial theorem and common formation/evolution processes (e.g. Busarello et al. 1997, Pahre et al. 1998). If ULIRG (and LIRG) mergers are indeed 'ellipticals in formation', they ought to fall on or near the fundamental plane. Placing the young, infrared luminous merger systems in this space and comparing them to (old) ellipticals is tricky, however, because of the influence of extinction, incomplete relaxation and stellar evolution, all of which strongly affect $\mu_{eff}$ and make $r_{eff}$ uncertain, even if determined in the near-infrared (see section 3.1).

Keeping these caveats in mind, the four insets in Fig.10 and Fig.11 show that *most ULIRGs are remarkably close to or on the fundamental plane of dynamically hot galaxies, in excellent agreement with the thesis that they will evolve into elliptical galaxies*. This conclusion also holds if we only consider the less evolution sensitive, $\sigma$-$r_{eff}$ projection of the fundamental plane (Fig.11 (top)). Only the ULIRG mergers with the smallest $r_{eff}$ (Arp 220, IRAS 01388-4618, NGC 6240, IRAS 00456-2901) are significantly offset from the fundamental plane (Fig.10). A similar conclusion holds for the LIRG mergers. The close association of ULIRGs with the fundamental plane of old ellipticals is actually quite surprising. All ULIRGs of our sample are actively forming stars at a high rate ($>10^2$ $M_\odot$ yr$^{-1}$), and in several of them deep K-band CO features indicate that red supergiants (ages 10-50 Myrs) or AGB stars (ages 100-600 Myrs) contribute significantly to their near-infrared light distribution (Goldader et al. 1995, Armus et al. 1995, Tecza et al. 2000). Even if the average stellar population were no younger the typical dynamical ages of the merger as estimated from the lengths of the tidal tails (a few hundred Myrs), their near-infrared emission should be 2 to 3.3 mag brighter than that of old ellipticals. A brightening of about 4 mag would be expected for an even younger supergiant population. The observed offset of ULIRGs (and LIRGs) with the smallest $r_{eff}$ in Fig.10 (bottom) and in the $r_{eff}$-$\mu_{eff}$ projection does not indicate more than ~1.4 mag of such an evolutionary brightening. Extinction can probably explain much or all of that difference. The coincidence of most ULIRGs with the fundamental plane thus is likely due to extinction and population effects partially canceling each other. An average extinction correction of A(K)~1 to 2.5 mag (or A(V)~11 to 28 mag) would account for the observed location and expected evolutionary brightening. Such extinction values are larger than estimated from the near-infrared colors or near-IR/optical hydrogen recombination line ratios, assuming a model of a foreground screen of absorbing dust (Goldader et al. 1995, Armus et al. 1995, Genzel et al. 1998, Scoville et al. 2000). They are fully consistent, however, with (more realistic) models where the absorbing dust and the stellar emission are mixed (Genzel et al. 1998). Lower extinction values are possible, if a significant fraction of the near-infrared light comes from stars much older than a few hundred Myrs.

The location along the fundamental plane is correlated with the luminosity (mass) and the dynamical/structural properties of ellipticals. Luminous (>>$L_*$, corresponding to $M(K)_*$~-24.3 and $M(B)_*$~-20.4), slowly rotating and anisotropic, giant ellipticals with boxy isophotes and large cores are found in the upper right of the fundamental plane (Fig.11). Somewhat lower luminosity (~$L_*$), faster rotating ellipticals and lenticulars with disky isophotes are found in the central part of the plane. Dwarf ellipticals are found to the lower left of the intermediate mass ellipticals. ULIRG (and LIRG) mergers match the locations of intermediate mass, disky ellipticals and S0s. ULIRGs have much smaller effective radii ($r_{eff}$< a few kpc, as compared to ≥10 kpc) and smaller velocity dispersions than giant ellipticals with boxy isophotal shapes. The one significant outlier again is NGC 6240, which lies in a poorly populated region of relatively small galaxies with very large velocity dispersions. In the fundamental plane NGC 6240 does not appear to be a giant elliptical in formation.

*3.6 ULIRG mergers rotate like disky ellipticals/lenticulars*

Fig.12 shows the $\sigma$-$r_{eff}$-$v_{rot}/\sigma$ distribution of ULIRGs and ellipticals/lenticulars. Hot galaxies with significant rotational support are located in the upper left (= disky ellipticals, S0s, bulges). Pressure supported (anisotropic) systems (=boxy ellipticals) are found in the lower right. Again ULIRGs better match the distribution of rotating (=disky) ellipticals and S0s in this diagram, and rotate faster than most boxy ellipticals with large cores. Pressure dominates ($v_{rot}/\sigma<1.7$) in all but the rapidly rotating, double nuclei systems Arp220 and NGC 6240.

*3.7 Importance of gas*

Downes and Solomon (1998) have observed with high-resolution millimeter interferometry four of the ULIRGs in the present sample: Arp 220, Mkn 273, IRAS 17208-0014 and IRAS 23365+3604. They conclude that molecular interstellar gas makes up 10 to 40% of the central dynamical mass/matter surface density. In NGC 6240 Tacconi et al (1999) estimate that the gas fraction could be 30%-70% of the dynamical mass. If one adds to this the fraction of mass in young stars, these two components make up a significant fraction of the total central mass densities of these galaxies. As such, the dissipative gas component, plus massive central bulges of the parent galaxies (which have a high density to start with) may thus be the key factors in overcoming the phase space constraints in forming the dense inner parts of ellipticals from disk galaxies, in accordance with the proposals of Kormendy and Sanders (1992).

# 4. DISCUSSION AND CONCLUSIONS

## *4.1 Confirmation of the 'ellipticals-in-formation' scenario*

Our infrared observations confirm on solid spectroscopic grounds that ULIRG mergers indeed appear to be forming moderately massive ($L_*$), field ellipticals. This conclusion is based on,

1. late stage ULIRG mergers fall on or near the fundamental plane of hot galaxies, and especially its less evolution sensitive, $r_{eff}$-$\sigma$ projection. Their stellar dynamics is largely relaxed. Pressure dominates but significant stellar rotation exists for most of our sample. Gas and stellar dynamics are decoupled, and

2. the distributions of velocity dispersions and effective radii, and the ratio of rotation velocity to dispersion in the ULIRG sample resemble those of intermediate mass, (~$L_*$), rotating (compact cores, disky isophotes) ellipticals and lenticulars. The ULIRGs in our sample are different from very luminous (massive), slowly rotating (large cores, boxy isophotes) giant ellipticals.

From these findings we conclude:

1. ULIRG mergers are the high luminosity, high mass tail of the possible range of present day and gas rich, disk galaxy mergers,

2. the end products of ULIRG mergers are intermediate mass (~$L_*$) elliptical galaxies, and

3. intermediate mass, field ellipticals can have plausibly formed through (recent) merging of gas rich, massive disk galaxies in the field, or in small groups. Giant ellipticals must have had a different formation history. Their common occurrence near the cores of galaxy clusters suggests that early, multiple and gas-poor mergers may have played a role.

*4.2 Comparison to numerical simulations*

Bendo and Barnes (2000) and Naab and Burkert (2001a) have recently studied the stellar dynamics of merger remnants with numerical simulations. They investigated different cases of orbital geometries for equal mass, 2:1, 3:1 and 4:1 mass ratio mergers, in total spanning a reasonably wide range of the likely phase space of parabolic encounters. They then 'observed' the resulting merger remnants after final coalescence. The 'artificial' slit spectra and maps of the kinematic moments of the numerical studies can be directly compared with our observations.

For the equal mass merger simulations the common features of the merger remnants are fairly complete relaxation with a modest amount of rotation ($v_{rot}/\sigma \sim \leq 0.2$ at $r_{eff}$), an approximately constant (in space) velocity dispersion within 1-2 times $r_{eff}$ and a wide range of 'abnormal' kinematic features, such as counter-rotation at large radii, orthogonally rotating cores and misaligned rotation axes. These 'abnormal' features depend on the initial orientations and spins of the progenitor disks. Equal mass mergers completely erase the memory of the initial disks.

In contrast unequal mass mergers are less violent and the more massive progenitor galaxy can survive relatively unscathed. The end product is typically an oblate, fast rotating remnant with $<v_{rot}/\sigma> \sim$ 0.5, 0.7 and 0.9 at $r \sim r_{eff}$ (for 2:1, 3:1 and 4:1 ratios). There are no strong kinematic misalignments. While the rotation curve of the remnant increases smoothly with radius and then levels off at $\sim 2r_{eff}$, the dispersion remains approximately constant within $r_{eff}$, and then drops rapidly at larger radii, where the initial disk of the more massive galaxy has survived without a great deal of dynamical heating. Naab and Burkert (2001a) conclude that equal mass mergers are the precursors of slowly rotating, boxy systems while unequal mass mergers evolve to rotating, disky systems.

The stellar dynamics observed in our late stage ULIRG mergers is fairly well matched by the results of the unequal mass merger simulations, especially when taking into account that our rotation velocities (Table 2, Fig.12) are lower limits and taken at $r \sim 0.5\text{-}1 \times r_{eff}$. On the other hand in our three ULIRGs with double nuclei (Arp 220, NGC 6240 and IRAS 14348-1447) the masses of the individual nuclei, as estimated from velocity dispersion and rotation, are remarkably similar (Table 2, Tecza et al. 2000, Sakamoto et al. 1999). Our study cannot address the issue of higher order kinematic moments and of 'abnormal' kinematic features and misalignments. The off-nuclear signal-to-noise ratio in most galaxies is not sufficient to determine more than $v_{rot}$ and $\sigma$. To explore the spatial distribution of the kinematic properties two dimensional data are required, which will become available in the near future with adaptive optics assisted, near-IR integral field spectrometers, such as SINFONI on the ESO VLT (Thatte et al. 1998).

*4.3 Evidence for recent star formation in disky ellipticals*

If (some) ellipticals have undergone recent major merger/starburst phases there should be evidence for young(er) stars in their optical spectra. The conventional view is, however, that luminous elliptical galaxies are old, coeval and were created about 15 Gyrs ago (e.g. Renzini 1999). The tight color-magnitude and $Mg_2$-$\sigma$ relationships and the small scatter of the fundamental plane lead to the interpretation that the stars making up the early type galaxies over a wide range of properties have all been formed at high redshift, independent of environment (e.g. Bender, Burstein & Faber 1993, Bernardi et al. 1998). Recent observations and population modeling have challenged this conventional interpretation of the data. Broad-band colors and $Mg_2$-indices are degenerate in metallicity and age, and it is possible that at fixed $\sigma$ young ages are offset by high metallicities, thus exactly preserving the $Mg_2$-$\sigma$ relationship (Trager et al. 2000a,b). Using additional stellar indices (H$\beta$ and Fe) and population synthesis models Trager et al. (2000a,b, and references therein) have been able to lift the age/metallicity degeneracy. They find that cluster ellipticals in Fornax and Virgo indeed are old and coeval. Ellipticals in lower density environments, however, span a much wider age range (1.5-18 Gyr), with a tendency for the somewhat lower $\sigma$ galaxies to be younger. Kuntschner and Davies (1998) also find that all Fornax ellipticals are old and coeval but that Fornax lenticular galaxies exhibit a substantial age spread. It thus appears that ellipticals in lower density environments, lower mass ellipticals and lenticular galaxies all exhibit evidence for relatively recent, substantial star formation episodes. This result is qualitatively consistent with a merger hypothesis and with our conclusions from the (U)LIRG merger samples.

*Acknowledgements.* We thank the staffs of the ESO VLT and of Keck for their excellent support during and after the observations, Ralf Bender for useful discussions, and Ric Davies for his invaluable help with the data reduction and access to his imaging data of IRAS 23365+3604. We thank Nick Scoville for making available to us the surface brightness distributions and most recent analysis of the NICMOS sample of ULIRGs, as well as Sylvain Veilleux and Dave Sanders for making available to us their unpublished surface brightness distributions of 1 Jy sample ULIRGs. We also acknowledge help with software and calibration details from Lowell Tacconi-Garman and Niranjan Thatte. We are grateful for useful comments by the referee, which helped to improve this paper.

# Appendix A: Comments on Individual Sources

IRAS 01388-4618. This source is single but, together with IRAS17208-0014, stands out because of its large gaseous and stellar, rotational component (Figure 4a, 8a). The gas disk is more extended than the stellar near-IR emission (Fig.8a) and shows a constant rotation velocity.

IRAS 01572+0009. This source (=Mkn 1014) is the most distant of our sample (z=0.163) and at the same time the only AGN/QSO powered system in our sample. Because of its redshift, the strong $\Delta v=3$ CO bands are not accessible (Figure 1). In addition strong hot dust emission from the QSO dilutes the stellar emission from the galaxy itself (Figure 4a). A 2.2 hour integration (two slits positions) concentrated on a fairly weak double absorption feature at rest wavelength ~1.50μm. The absorption feature is detected but the achieved signal to noise ratio is marginal for a good determination of velocity dispersion (Figure 6). No dip is seen in the absorption depth between the two features, implying that the velocity dispersion is greater than 170 km/s. In our further analysis we adopted a velocity dispersion of 200 km/s that is a conservative estimate but has a large uncertainty (~70 km/s).

IRAS 14348-1447. The NICMOS image of this double nucleus system (separation 5.3 kpc) shows two similar size disks with spiral arms and a ~Sb nucleus/disk morphology (Scoville et al. 2000). While the southern nucleus is somewhat brighter (factor 2) and has somewhat greater velocity dispersion (170 vs. 150 km/s), the two nuclei nevertheless are remarkably similar in their dynamical properties/masses, in agreement with our findings for NGC 6240 and Arp 220. Both galaxies appear to have small inclinations (~0-20° for the more massive southern and 25-30° for the northern galaxy). The small inclinations deduced from the near-IR morphology are in agreement with their small observed, gas and stellar rotation velocities (~45 and ~70 km/s, Figure 8a, Table 2). IRAS 14348-1447 appears to be a prograde-prograde merger.

Arp 220. Arp 220 is the nearest ULIRG ($D_L$=77 Mpc). It has two near-IR 'nuclei' at a separation of ~0.9" (~330 pc, Scoville et al. 1998). Our new Arp220 stellar dynamics (Figures 4b and 7) is similar but not identical to the sub-arcsecond, gas dynamics obtained by Downes and Solomon (1998) and Sakamoto et al. (1999) with J=2-1 CO 1.3mm millimeter interferometry. The near-IR and millimeter data have comparable spatial resolution. The differences between near-IR stellar and millimeter gas dynamics cannot be caused by the substantial near-IR extinction ($A_K$(screen)≥2 mag) since the v=2-1 S(1) line emission of $H_2$ (obtained at the same wavelength and resolution as the stellar data) exhibits exactly the same kinematics as the millimeter line. We confirm the large velocity gradient along position angle 52° through the eastern emission nucleus (Figures 7, 8b) that was found in the millimeter data. An east-west slit through both nuclei shows similar but opposite velocity gradients (Figures 7, 8b). Our data are in excellent agreement with the interpretation of Sakamoto et al. (1999) that the two near-IR peaks contain the progenitor nuclei of the merger and consist of young stars, gas and a substantial mass of older stars. They have large rotation velocities, and hence large masses (≥2x10$^9$ $M_\odot$ within r=110 pc). If this interpretation is adopted the two nuclei are in counter-rotation, as proposed by Sakamoto et al. (1999). Eckart and Downes (2001) have recently proposed the interesting alternative interpretation that the inner region of Arp 220 is a highly warped, rotating disk, centered between the eastern and western near-IR nuclei. In this scenario, the two near-IR and millimeter peaks are caused by line-of-sight crowding of the warped disk, which is seen at large inclination angles in these two directions. The apparent counter-rotation would be a natural consequence of the symmetry of the disk. While this model can plausibly account for the generic properties of the data, it needs to be tuned to account for the measurements more quantitatively, and fails to explain the difference between gas and stellar dynamics shown in Figure 7. It can also not easily explain the significant difference in flux and small scale morphology of the two near-IR peaks. Finally the nature of the central mass between the two emission peaks is left unexplained. Unlike in the case of NGC 6240 (peak of velocity dispersion between nuclei = peak of molecular gas density) there is no gas concentration between the two nuclei, or any other indication for a mass concentration. Given that many other ULIRGs (e.g. NGC 6240 and IRAS 14348-1447, Scoville et al. 2000, Figures 4b) also have double near-IR peaks, but in addition, have clearly recognizable individual bulges and disks, we find the counter-rotating, double nucleus interpretation to be the much more plausible one. The similar rotation amplitudes and velocity dispersion of the two nuclei indicate that the two bulges are of very similar mass.

NGC6240. Our new H-band stellar data are in excellent agreement with the earlier K-band imaging spectroscopy obtained with the MPE 3D spectrometer by Tecza et al. (2000). Both nuclei counter-rotate in a retrograde-prograde overall geometry. The rotation velocities of the two nuclei are of similar size and are very large (~300±70 km/s, Fig.8 of Tecza et al. 2000). The velocity dispersion is large throughout the region encompassing both nuclei but has a maximum between the nuclei, on/near the millimeter CO emission peak found with 1.3 millimeter interferometry by Tacconi et al. (1999). The large velocity dispersion at the CO peak cannot be explained by the rotation properties of the two disks that may overlap in that direction. The line connecting the two galaxies (and passing through the CO peak) is at large angles relative to the line of nodes of both galaxy bulges. Hence velocity gradients along this line are fairly small and the velocity dispersion peak must be due to a local effect. Our LOSVDs in Figure 5 suggest that the large velocity dispersion at the CO peak is caused by a blue-shifted component (300-600 km/s) that is not present in the other directions, again in excellent agreement with the Tecza et al. (2000) K-band data. We thus concur with Tacconi et al. (1999) and Tecza et al. (2000) that there probably is a major self-gravitating mass concentration of interstellar gas between the two nuclei, although the possibility of non-equilibrium dynamics has to be taken into account (Mihos 1999).

IRAS 17208-0014. The NICMOS image of this object shows a compact, single nuclear source elongated along position angle 90-110$^o$, surrounded by a totally disrupted disk. Stars and gas are rotating rapidly and with similar properties (line of nodes, rotation velocity). The kinematic properties deduced from our near-IR data (Figures 4b, 8a) are in good agreement with those obtained from CO millimeter interferometry (Downes and Solomon 1998). This source has the second largest overall velocity dispersion after NGC 6240. The [FeII] position-velocity diagram indicates that the gas disk is more extended than the near-IR stellar emission and shows a constant rotation velocity beyond a bright inner ring.

IRAS 23365+3604. In this system the gas and stellar dynamics are strongly decoupled. The gas emission comes from a fairly extended rotating disk with a position angle of the lines of nodes at -30 to -45°. This rotating disk is seen in P$\alpha$ in our data (Figures 4c, 8a) but also in CO millimeter emission in the interferometric data of Downes and Solomon (1998). The modeling of Downes and Solomon (1998) suggests a small (30°) inclination, implying a fairly large inclination correction to the observed (~60 km/s) rotation velocity (Figure 8a). In contrast the stellar data show very little rotation and the stellar dynamics is relaxed.

IRAS 23578-5307. This source is the second most distant and only fairly faint features near rest-wavelength 1.56µm are accessible at the long wavelength side of the H-band (see Figure 1). The source is a moderate luminosity ULIRG only, implying that it is faint. Our integrated spectrum (two slits averaged, 1.7 hour integration, Figure 4c) has a fairly low signal-to-noise ratio, and we can only make an uncertain (±70 km/s) determination of the dispersion.

*Table 1*
*Source List*

| source | R.A.(2000) | Dec.(2000) | z | Scale (kpc/arcsec) | log ($L_{IR}$) | p.a. slits | Integr. time (hours) | Instrum. |
|---|---|---|---|---|---|---|---|---|
| 00262+4251 | $00^h28^m54.0^s$ | +43°08'18" | 0.0927 | 1.74 | 12.0 | +45°, 0° | 0.3, 1 | NIRSP |
| 00456-2901 | 00 48 03.6 | -28 48 38 | 0.1103 | 1.94 | 12.2 | +30 | 0.7 | ISAAC |
| 01388-4618 | 01 40 55.9 | -46 02 53 | 0.0903 | 1.65 | 12.0 | 0, +90 | 0.7, 0.7 | ISAAC |
| 01572+0009 (Mkn 1014) | 01 59 50.2 | +00 23 41 | 0.1630 | 2.67 | 12.5 | +20, -70 | 1.2, 1 | ISAAC, NIRSP |
| 14348-1447 | 14 37 38.3 | -15 00 23 | 0.0823 | 1.51 | 12.3 | +30 | 4 | ISAAC |
| 15327+2340 (Arp 220) | 15 34 57.1 | +23 30 11 | 0.0181 | 0.36 | 12.1 | +52, -91 | 2, 2 | ISAAC |
| 16504+0228 (NGC 6240) | 16 52 58.9 | +02 24 03 | 0.0245 | 0.50 | 11.8 | -158, -31 | 0.3, 0.3 | ISAAC |
| 17208-0014 | 17 23 21.9 | -00 17 00 | 0.0428 | 0.83 | 12.3 | +90, +120 | 0.5, 0.5 | ISAAC |
| 20087-0308 | 20 11 23.2 | -02 59 54 | 0.1057 | 1.88 | 12.4 | -45, +45 | 0.7, 0.7 | ISAAC |
| 20551-4250 | 20 58 26.9 | -42 39 06 | 0.0428 | 0.85 | 12.0 | -45, +45 | 1, 1 | ISAAC |
| 23365+3604 | 23 39 01.3 | +36 21 10 | 0.0645 | 1.21 | 12.1 | +45, -30 | 0.2, 0.7 | NIRSP |
| 23578-5307 | 00 00 23.6 | -52 50 28 | 0.1249 | 2.20 | 12.1 | +107, 14 | 1, 0.7 | ISAAC |

*TABLE 2*
*Dynamical/Structural Properties of ULIRG Mergers[a]*

| Galaxy | z | $\sigma$ stars | $v_{rot}$ stars | $v_{rot}/\sigma$ stars | $v_{rot}$ gas | $v_{gas}/v_{stars}$ | $r_{eff}$ kpc | $\mu_{eff}(K)$ (mag) | $M(K)_{tot}$ (mag) | notes |
|---|---|---|---|---|---|---|---|---|---|---|
| I 00262+4251 | 0.0970 | 170(15) | <15(10) | <0.09 | <60(15) | ~4 | 3.4(1) | | | b |
| I 00456-2901 | 0.1099 | 162(25) | 45(10) | 0.28 | | | 1.3(0.5) | 15.8 | -25.8 | c |
| I 01388-4618 | 0.0912 | 144(10) | 130(15) | 0.90 | 130(30) | 1.00 | 1.0(0.4) | 13.7 | -25.2 | d |
| I 01572+0009 | 0.1630 | 200(60) | | | | | 4.0(2) | 16.5 | -26.2 | e |
| I 14348-1447u | 0.0824 | 150(25) | 50(15) | 0.33 | 87(10) | 1.74 | 3(1) | 17 | -24.7 | f |
| I 14348-1447l | 0.0829 | 170(14) | 60(20) | 0.35 | 30(20) | 0.50 | 3(1) | 17 | -24.7 | f,g |
| Arp220(W) | 0.0180 | 172(10) | >90(30) | >0.52 | yes | | 0.6(0.3) | 13.9 | -24.0 | h |
| Arp220(E) | 0.0182 | 155(10) | 185(30) | 1.19 | 185(50) | 1.00 | 0.6(0.3) | 13.9 | -24.0 | h,i |
| NGC 6240(N) | 0.0249 | 270(25) | 210(40) | 0.78 | yes | | 1.1(0.3) | 13.7 | -25.5 | j |
| NGC 6240(S) | 0.0251 | 290(25) | 280(50) | 0.97 | yes | | 1.1(0.3) | 13.7 | -25.5 | j,k |
| I 17208-0014 | 0.0429 | 229(15) | 110(20) | 0.48 | 170(20) | 1.52 | 1.4(0.3) | 14.6 | -25.3 | l |
| I 20087-0308 | 0.1057 | 219(14) | 50(15) | 0.23 | | | 3.3(1) | 15.9 | -25.8 | m |
| I 20551-4250 | 0.0431 | 140(15) | 40(10) | 0.29 | 57(7) | 1.43 | 2.1(0.7) | 16.4 | -24.0 | n |
| I 23365+3604 | 0.0644 | 145(15) | <15(10) | <0.1 | 55(5) | >3.7 | 4.8(1) | 17 | -25.5 | o |
| I 23578-5307 | 0.1250 | 190(70) | | | | | 9.1(5) | 18.1 | -25.7 | p |
| Mkn 273 | 0.0371 | 160(60) | | | | | 1.3(0.4) | 14.5 | -25.1 | q |

Notes to Table 2:

a) The (central) velocity dispersions are typically in an aperture of 0.6"x (1-1.5"). Rotation velocities are half of the intensity weighted, velocity differences on either side of the galaxy, at radii of 1-2". They do not include corrections for inclination and beam smearing and represent lower limits to the true rotation speed. Numbers in parentheses are the uncertainties ($\pm$), including systematic effects. The uncertainty of $\mu_{eff}$ ($r \leq r_{eff}$) (K) is about 0.7 mag, $M(K)_{tot}$ is the absolute magnitude of the entire merger system, $M(K)_* \sim -24.3$
b) FWHM gas 400-500 km/s, $r_{eff}$ from Keck H-band images
c) $r_{eff}$ and $\mu_{eff}$ from K-band data of Veilleux et al.(2001)
d) rotation along both slits, $r_{eff}$ from VLT H-band data (Fig.3) and calibration for $\mu_{eff}$(K) from Rigopoulou et al. (2000)
e) $r_{eff}$ is quite uncertain due to AGN contribution (Fig.3), $r_{eff}$ is an average of VLT H-band data (Fig.3), the K-band data of Scoville et al. (2000) and the K-band data of Veilleux et al. (2001)
f) total velocity dispersion of both nuclei 165 km/s; pro-grade, pro-grade merger, $r_{eff}$ and $\mu_{eff}$ from the K-band data of Veilleux et al. (2001) at r>9 kpc, outside the two nuclei (separation ~4"=6 kpc)
g) total velocity dispersion of both nuclei 165 km/s; very compact nucleus
h) total velocity dispersion of both nuclei 165 km/s; pro-grade, retro-grade merger. Our data do not provide a good gas rotation velocity for the western nucleus because the E-W slit is not along the line of nodes of the gas rotation. The Sakamoto et al. (1999) CO 2-1 data give $(v_{rot}(gas)sini)_{peak}$ =120 km/s for the western nucleus. $r_{eff}$ and $\mu_{eff}$ are from the K-band data of Scoville et al. (2000); correction for the K-band extinction ($\geq 1$ mag) decreases $r_{eff}$ to ~0.4 kpc.
i) total velocity dispersion of both nuclei is 165 km/s. The Sakamoto et al. (1999) CO 2-1 data give $(v_{rot}(gas)sini)_{peak}$ =160 km/s for the eastern nucleus.
j) total velocity dispersion of both nuclei 320 km/s (rotation not subtracted); pro-grade, retro-grade merger. The molecular gas does exhibit rotation but its magnitude and direction are very uncertain owing to the large random motions (Tacconi et al. 1999, Tecza et al. 2000). $r_{eff}$ and $\mu_{eff}$ are from the K-band data of Scoville et al. (2000).
k) total velocity dispersion of both nuclei 320 km/s (rotation not subtracted). The molecular gas does exhibit rotation but its magnitude and direction are very uncertain owing to the large random motions (Tacconi et al. 1999, Tecza et al. 2000).
l) rotation along both slits. $r_{eff}$ and $\mu_{eff}$ are from the K-band data of Scoville et al. (2000).
m) rotation seen along slit 2. $r_{eff}$ is from the VLT H-band data (Fig.3). $\mu_{eff}$ is from the H-band data calibrated with the K-band magnitude of Duc et al. (1997)
n) rotation along both slits. $r_{eff}$ is from the VLT H-band data (Fig.3). $\mu_{eff}$ is from the H-band data calibrated with the K-band magnitude of Duc et al. (1997)
o) $r_{eff}$ is from the H-band data of Davies et al. (2001, Fig.3). $\mu_{eff}$ is from the H-band data calibrated with the K-band magnitude of Surace et al. (2000)
p) $r_{eff}$ is from the VLT H-band data. $\mu_{eff}$ is from the H-band data calibrated with the K-band magnitude of Rigopoulou et al. (2000)
q) data from James et al. (1999)

*TABLE 3*
*Dynamical/Structural Properties of LIRG Mergers*[a]

| Galaxy | σ stars | $r_{eff}$ kpc | $\mu_{eff}(K)$ (mag) |
|---|---|---|---|
| NGC 3690 B2 | 66(23) | 0.22(0.05) | 14.3 |
| IC 694 | 135(47) | 0.36(0.08) | 14.2 |
| NGC2623 | 95(33) | 0.37(0.08) | 14.6 |
| NGC 1614 | 76(26) | 0.5(0.11) | 14.0 |
| I 17138-1017n | 72(25) | 0.54(0.12) | 14.8 |
| I 17138-1017s | 104(36) | 0.54(0.12) | 15.1 |
| Zw475.056 | 151(50) | 0.65(0.15) | 15.5 |
| Mkn 331 | 101(35) | 0.65(0.15) | 12.2 |
| MCG+05-6-36 | 138(40) | 1.73(0.4) | 14.5 |
| IC883 | 206(90) | 3.9(0.45) | 16.4 |
| IC 4553 | 150(21) | 4(0.3) | 16.1 |
| NGC 4194 | 104(25) | 1.1(0.1) | 14.7 |
| NGC 6090 | 50(20) | 5.4(0.4) | 16.5 |
| NGC 7252 | 123(19) | 3.9(0.2) | 15.9 |

Notes to Table 3:

a) Data are from Shier and Fischer (1998) and James et al. (1999). Numbers in parentheses are the uncertainties (±). The uncertainty of $\mu_{eff}$ ($r \leq r_{eff}$) (K) is about 0.5 mag

*Table 4*
*Statistical Properties of Samples*

| Sample | Number | Mean velocity dispersion (km/s) | Median velocity dispersion (km/s) | Uncertainty in mean/median (km/s) | Reference |
|---|---|---|---|---|---|
| ULIRGs | 13 | 185 | 165 | 14 | this paper, James et al. 1999 |
| LIRGs | 12 | 111 | 103 | 13 | James et al. 1999, Shier and Fischer 1999 |
| Merger remnants | 13 | 201 | 186 | 13 | Lake and Dressler 1986 |
| Giant Boxy ellipticals ($M_B \leq -20.5$) | 36 | 269 | 285 | 9 | Faber et al. 97, Bender et al. 1992 |
| Intermediate mass, disky ellipticals and S0s ($-18.5 \leq M_B \leq -20.5$) | 27 | 166 | 164 | 7 | Faber et al. 97, Bender et al. 1992 |
| Sa-c spiral bulges | 94 | 146 | 139 | 5 | Whitmore et al. 85, Bender et al. 1992 |
| Sa-b spiral bulges | 63 | 151 | 156 | 6 | Whitmore et al. 85, Bender et al. 1992 |

# Figure Captions

Figure 1: Atmospheric transmission and stellar features in the H-band. The upper curve is the transmission of the Earth's atmosphere at airmass 1 as calculated from the ATRAN code (S.Lord, priv.comm.). The lower curve is the NIRSPEC ($\lambda/\Delta\lambda=2200$) spectrum of the K5 Ib supergiant HD200576 (normalized and shifted by –0.45). Vertical bars mark the position of 1.6μm rest wavelength at different redshifts. The strong H-band CO bandheads can be observed in the H-band out to about z~0.12.

Figure 2: Stellar template spectra. Left: M0III giant HD25472 used as template for the $\lambda/\Delta\lambda=5200$ H-band ISAAC spectroscopy at the VLT. Upper right: Comparison of the NIRSPEC H-band template HD200576 (K5 Ib, see also Fig.1, short dashes) with the K5 III giant HD227277 (long dashes, also NIRSPEC) and the M0III ISAAC template (smoothed to the same R=2200 resolving power). Bottom right: K-band template HD99817 (M1 Iab) used for the Arp220 ISAAC VLT observations.

Figure 3: H-band surface brightness as a function of (major axis radius)$^{1/4}$, for 6 of the program galaxies. The continuous curves are the surface brightness data as determined from ellipse fits to the H-band images in the IRAF/STSDAS routine 'Ellipse'. The dashed curves are the best de Vaucouleurs law ($r^{1/4}$) fits to the data, away from the very center determined by seeing or a central point source (IRAS 01572+0009). The best fit, effective (=half energy) radii are listed for each galaxy. The central surface brightness values $I_0$ are: $3.3 \times 10^{-3}$ Jy/arcsec$^2$ for IRAS01388-4618, as calibrated on the total H-band flux density of $8.5 \times 10^{-3}$ Jy from Rigopoulou et al. (2000), $8.2 \times 10^{-3}$ Jy/arcsec$^2$ for IRAS01572+0009 (Mkn 1014), as calibrated on the total H-band flux density of $5.7 \times 10^{-3}$ Jy of Scoville et al. (2000), $1.9 \times 10^{-3}$ Jy/arcsec$^2$ for IRAS20087-0308, as calibrated on the total H-band flux density of $3.8 \times 10^{-3}$ Jy of Duc et al. (1997), $8.4 \times 10^{-4}$ Jy/arcsec$^2$ for IRAS20551-4250, as calibrated on the total H-band flux density of $1.04 \times 10^{-2}$ Jy of Duc et al. (1997), and $2 \times 10^{-3}$ Jy/arcsec$^2$ for IRAS23365+3604 (data from Davies et al. 2001), as calibrated on the total H-band flux density of $8.9 \times 10^{-3}$ Jy of Surace et al. (2000).

Figure 4: Integrated source spectra. Figure 4a. Top left: source integrated NIRSPEC ($\lambda/\Delta\lambda \sim 2200$) H/K-band spectra (normalized) along position angles $0^o$ (top), and $45^o$ (bottom) of the z=0.098 ULIRG IRAS 00262+4251, as a function of rest-wavelength. The noisy part of the spectrum between rest-wavelengths 1.67 and 1.75µm is caused by poor transmission of the Earth's atmosphere between the H and K-bands. Identifications of several emission lines are given. Top right: source integrated ISAAC ($\lambda/\Delta\lambda \sim 5200$) H-band spectrum (normalized and slope removed) of the z=0.11 ULIRG IRAS 00456-2901 as a function of rest-wavelength. Bottom left: source integrated (both slits) ISAAC ($\lambda/\Delta\lambda \sim 5200$) H-band spectrum (normalized and slope removed) of the z=0.09 ULIRG IRAS 01388-4618, as a function of rest-wavelength. Bottom right: source integrated, normalized NIRSPEC ($\lambda/\Delta\lambda \sim 2200$) H/K-band spectrum of the z=0.16 ULIRG IRAS 001572+0009 (Mkn 1014), as a function of rest-wavelength. The inset gives the integrated, normalized ISAAC ($\lambda/\Delta\lambda \sim 5200$) H-band spectrum near 1.5µm.

Figure 4b. Top left: source integrated, normalized (and slope removed) ISAAC ($\lambda/\Delta\lambda \sim 5200$) H-band spectra of the SW (upper trace) and NE (lower trace) nuclei of the z=0.08 ULIRG IRAS 14348-1447, as a function of rest-wavelength. Top right: source

integrated (0.9" along slit), normalized (and slope removed) ISAAC ($\lambda/\Delta\lambda$~5200) K-band spectra of the eastern (upper) and western (lower) nuclei of Arp 220 (z=0.02).

Bottom left: Normalized (and slope removed) ISAAC ($\lambda/\Delta\lambda$~5200) H-band spectra of the southern nucleus (bottom trace, aperture 0.6"x0.6"), the northern nucleus (top trace, 0.6"x1.1") and the CO 2-1 emission peak between the two nuclei (middle trace, 0.6"x1.1") in the z=0.025 (U)LIRG NGC 6240, as a function of rest-wavelength. Bottom right: Normalized (and slope removed) ISAAC ($\lambda/\Delta\lambda$~5200) H-band spectra of ~1" apertures on the nucleus (middle trace), 1"W (bottom trace) and 1" E (top trace) of the nucleus of the z=0.04 ULIRG IRAS 17208-0014, as a function of rest wavelength.

Figure 4c. Top left: source integrated (both slits) ISAAC ($\lambda/\Delta\lambda$~5200) H-band spectrum (normalized and slope removed) of the z=0.11 ULIRG IRAS 20087-0308, as a function of rest-wavelength. Top right: source integrated, normalized (and slope removed) ISAAC ($\lambda/\Delta\lambda$~5200) H-band spectrum (both slits) of the z=0.04 ULIRG IRAS 20551-4250, as a function of rest-wavelength. Bottom left: source integrated, normalized (and slope removed) NIRSPEC ($\lambda/\Delta\lambda$~2200) H/K-band spectrum (position angle –30$^o$) of the z=0.06 ULIRG IRAS23365+3604, as a function of rest-wavelength. Bottom right: source integrated ISAAC ($\lambda/\Delta\lambda$~5200) H-band spectrum (normalized and slope removed) of the z=0.13 ULIRG IRAS 23578-5307, as a function of rest-wavelength.

Figure 5. Correlation amplitude as a function of velocity offset from best-fit systemic velocity for 10 program galaxies. The solid curves are the outputs of the Fourier-quotient, cross correlation analysis described in the text. The dashed curve is the best fitting Gaussian to the cross correlation amplitude, with the dispersion (and systematic)

uncertainty listed for each galaxy. For the VLT ISAAC H-band data we used the template star HD25472 (M0 III), for the Keck NIRSPEC H-band data we used the star HD200576 (K5 Ib) and for the VLT ISAAC K-band data we used the star HD99817 (M1 Ia/b).

Figure 6. Comparison of source spectra (solid) with the best-fit dispersion convolved template star spectra (dashed) for all 12 program galaxies. For the double nuclei galaxies we show system averaged spectra.

Figure 7. Position-peak velocity and position-velocity dispersion plots of the stellar and gas components in Arp220. The left inset shows the results for position angle $52^o$ through the eastern nucleus, the right inset shows the results for an east-west slit through both nuclei. Filled circles give the derived peak stellar velocities, open rectangles the velocities of the 2.248 $\mu$m v=2-1 S(1) emission lines of $H_2$, and the thick continuous lines (without symbols) the velocities of the CO J=2-1 1.3 mm rotational emission line (from Sakamoto et al. 1999). The top trace is the velocity dispersion derived from the K-band stellar spectra.

Figure 8a: Position (up-down)-velocity (left right) images of gas emission lines in 6 different ULIRGs of our sample, taken with NIRSPEC ($\lambda/\Delta\lambda$~2200, top left and bottom right) and ISAAC ($\lambda/\Delta\lambda$~5200, for the remainder). Spatial and velocity scales are indicated. The spectra have been continuum subtracted.

Figure 8b: Position (up-down)-velocity (left right) images of the 2.248 μm v=2-1 S(1) emission line of $H_2$ in Arp220, observed with ISAAC (λ/Δλ~5200). The left inset shows an east-west slit through both nuclei, the right inset shows a position angle 52° slit through the eastern nucleus. Spatial and velocity scales are indicated. The spectra have been continuum subtracted.

Figure 9: Distribution of velocity dispersions of our ULIRG sample (plus Mkn 273, James et al. 1999), compared with different samples of early type galaxies and spiral bulges (see text). Spiral bulges are Sa-Sb bulges from Whitmore et al. (1985). Giant boxy ellipticals ($M_B$≤-20.5) and intermediate disky ellipticals (-18.5≥$M_B$≥-20.5) are from Faber et al. (1997) and Bender et al. (1992). For the double nuclei ULIRGs we took the source integrated velocity dispersions listed in Table 2. Error bars are statistical. The distributions are binned in widths of 25 km/s.

Figure 10. top (a) : $\sigma$-$r_{eff}$-$\mu_{eff}$(K) distributions of ULIRG mergers (filled red squares) and ellipticals from the compilation of Pahre (1999) (grey crosses). The fundamental plane can be clearly seen. Bottom (b): distribution of ULIRG mergers (filled red squares, with error bars) perpendicular to the fundamental plane. For comparison filled green circles (with error bars) denote the LIRG mergers from Shier and Fischer (1998) and James et al. (1999), the open grey circles indicate ellipticals/lenticulars from the K-band survey of Pahre (1999), and open magenta triangles indicate ellipticals in the Coma cluster from the K-survey of Mobasher et al. (1999). The dashed line is the best fit slope

of the fundamental plane of hot galaxies, as determined by the K-band studies of Mobasher et al. (1999) and Pahre (1999): log $r_{eff}$= 1.4 (log $\sigma$ + 0.22 $\mu_{eff}$ (K)) + const. Also indicated are the influences of different extinction values (bottom) and of age/evolution of the stellar population (top).

Figure 11. top (a) : $\sigma$-$r_{eff}$ projection of mergers and ellipticals/lenticulars. ULIRG mergers are shown as filled red squares, LIRG mergers from Shier and Fischer (1998) and James et al. (1999) are given as green filled circles, ellipticals from the compilation of Pahre (1999) are shown as grey crosses, boxy giant ellipticals (M(B)<-20.5, B-K=3.9) from Bender et al. (1992) and Faber et al. (1997) are light blue squares with crosses and disky ellipticals (M(B)<-18, B-K=3.9) from Bender et al. (1992) and Faber et al. (1997) are shown as dark blue, filled circles. Bottom (b): $r_{eff}$-$\mu_{eff}$ projection of mergers and ellipticals. Symbols are the same as in the top inset.

Figure 12. $\sigma$-$r_{eff}$-$v_{rot}$/$\sigma$ distribution of ULIRG mergers (filled red squares), boxy giant ellipticals (light blue, open squares with crosses, M(B)<-20.5, B-K=3.9) and disky ellipticals/lenticulars (dark blue, filled circles, M(B)<-18.5, B-K=3.9) (both from Bender et al. 1992, Faber et al. 1997). Note that because of beam smearing and inclination effects, the ULIRG rotation velocities are lower limits.